\documentclass[review,11pt]{elsarticle}
\usepackage{epstopdf} 
\usepackage[pagewise,mathlines]{lineno}

\usepackage[top=2.75cm,bottom=2.75cm,left=3.5cm,right=3.5cm]{geometry}

\newcommand{\tabincell}[2]{\begin{tabular}{@{}#1@{}}#2\end{tabular}}

\usepackage{amsmath}
\usepackage{bm}
\usepackage{threeparttable}
\usepackage{amssymb}
\usepackage{array}

\usepackage{graphicx}
\usepackage{subfigure}

\usepackage{color}

\begin{document}

\begin{frontmatter}
\title{Design of a non-linear human breast diffusion coil}

\author[rvt]{Feng Jia\corref{cor1}}
\ead{feng.jia@uniklinik-freiburg.de}
\author[rvt]{Sebastian Littin}
\author[rvt]{Philipp Amrein}
\author[rvt]{Stefan Kroboth}
\author[rvt]{Huijun Yu}
\author[rvb]{Arthur W. Magill}
\author[rvb]{Tristan Kuder}
\author[rvc]{Sebastian Bickelhaupt}
\author[rvd]{Frederik Laun}
\author[rvb]{Mark E. Ladd}
\author[rvt]{Maxim Zaitsev}

\cortext[cor1]{Corresponding author}

\address[rvt]{Dept.of Radiology, Medical Physics, Medical Center University of Freiburg, Faculty of Medicine, University of Freiburg, Freiburg, Germany}
\address[rvb]{Medical Physics in Radiology, German Cancer Research Center, Heidelberg, Germany}
\address[rvc]{Junior Group Medical Imaging and Radiology-Cancer Prevention, German Cancer Research Center, Heidelberg, Germany}
\address[rvd]{Department of Radiology, MR Physics, University Medical Center Erlangen, Erlangen, Germany}


\begin{abstract}

Diffusion-weighted imaging (DWI) in the female breast is a magnetic resonance imaging (MRI) technique, which may potentially replace or effectively complement both x-ray mammography and contrast-enhanced MR mammography. To further improve specificity of DWI in the breast, stronger and faster diffusion weighting is advantageous. Here, a dedicated gradient coil is designed, targeted at diffusion weighting in the female breast, with the peak gradient magnitude exceeding that of the current clinical MR scanners by an order of a magnitude. 

Design of application-tailored gradient coils in MRI has recently attracted increased attention. With the target application in mind, the gradient coil is designed on an irregularly shaped semi-open current-carrying surface. Due to the coil former closely fitting the non-spherical target region, non-linear encoding fields become particularly advantageous for achieving locally exceptionally high gradient strengths. As the breast tissue has a predominantly isotropic cellular microstructure, the direction of the diffusion-weighting gradient may be allowed to vary within the target volume. However, due to the quadratic dependency of the b-factor on the gradient strength, variation of the gradient magnitude needs to be carefully controlled.  

To achieve the above design goals the corresponding multi-objective optimization problem is reformulated as a constrained optimization, allowing for flexible and precise control of the coil properties. A novel constraint is proposed limiting the gradient magnitude variation within every slice while allowing for variations both the direction of the gradient within the slice and the magnitude across the slices. The above innovations enable the design of a unilateral coil for diffusion weighting in the female breast with the local gradient strengths exceeding 1~T/m with highly homogeneous diffusion weighting for imaging in the coronal slice orientation. 

\end{abstract}


\end{frontmatter}


\section{Introduction}
\label{sec:Introduction} 
Modern magnetic resonance imaging (MRI) scanners can yield impressive and clinically relevant images, but owing to the limited sensitivity, the resolution is limited to the millimetre range. Direct imaging of individual cells in vivo, which typically have diameters below 10~$\mu$m, is not possible. Images as obtained in histology are beyond the current sensitivity limit by several orders of magnitude. Several magnetic resonance (MR) techniques exist that allow one to gain information about the tissue microstructure beyond the actual resolution limit of the direct image encoding. The most prominent of these techniques is diffusion-weighted imaging (DWI)~\cite{Jones_2011}. In DWI, the signal is sensitized to diffusion by applying additional balanced magnetic gradient pulses before the imaging readout. DWI is currently used to assess the tissue microstructure by making use of, for example, the often observed negative correlation between cell density and measured apparent diffusion coefficient (ADC). It has been observed that the ADC is often decreased in tumorous tissue, which makes its depiction a valuable tool in cancer diagnostics, e.g. of the prostate~\cite{Kim_Diffusion_2010}, brain~\cite{Stieltjes_Diffusion_2006}, or breast~\cite{OFlynn_Diffusion_2014}. Moreover, DWI has been demonstrated to help to improve the differentiation between malignant and benign breast lesions~\cite{zhang_multiparametric_2019}, potentially saving patients from unnecessary invasive procedures such as biopsies.

This is of significant importance with regards to clinical care, since standard diagnostic imaging methods commonly suffer from a relatively low positive predictive value. In the German breast cancer screening alone about 31,000 women are undergoing invasive procedures such as biopsy in order to clarify a suspicious finding detected using X-ray mammography. However, in about 50\% of the cases, the invasive procedure turns out to demonstrate a benign finding – thus in retrospect has been an unnecessary procedure~\cite{kooperation_2020}.
Approaches to allow further non-invasive clarification are therefore needed, ideally without the necessity to apply intravenous gadolinium containing contrast agents, which are under discussion due to studies demonstrating deposition in the body and potential neuropathic effects demonstrated in animal studies~\cite{radbruch_is_2020}.

However, the coupling of the apparent diffusion parameters to the tissue microstructure is in fact very loose, which currently limits the specificity of DWI when reporting on the underlying biophysical and cellular processes. The main reason for choosing this approach in vivo is the lack of technical alternatives. The gradient amplitudes achievable on current clinical scanners are too low to generate a dephasing of the magnetization on the scale of the cell dimensions in a time short enough to freeze diffusion-induced displacements on that spatial scale. In order to gain more accurate information about the tissue microstructure in human breast in vivo with DWI, a high performance gradient coil would be a key component\cite{jones_microstructural_2018}. 

The highest performance whole-body gradient system with 300 mT/m gradient strength is currently installed in the experimental Connectome scanner~\cite{setsompop_pushing_2013}. For safe in vivo operation of this system, a number of concerns with peripheral, visual and direct cardiac stimulation had to be resolved, rendering it impossible to exploit the technically achievable combination of the maximum gradient strength and slew rate~\cite{setsompop_pushing_2013}. Another solution to obtain a high gradient strength is to build a local gradient coil specific for human breast. Currently, a local breast gradient coil~\cite{VeldenKlomp2016} with the gradient strength of 300 mT/m was constructed with a potential application in DWI. Similar to the above-mentioned whole-body gradients, this gradient coil was designed to generate a conventional linear encoding field.

Recently, a non-linear gradient coil design methodology~\cite{jia_design_2017} has been proposed to maximize the gradient strength when a target field is unknown, which has been successfully used to design coil elements for localized high resolution imaging applications such as PatLoc~\cite{HennigZaitsev2008a}. Moreover, a breast coil design with field non-linearity larger than 5\%~\cite{JiaZaitsevBreast2018} has been investigated. The results have demonstrated that coil efficiency $\eta$, defined as the ratio of gradient strength of the encoding field to the current flowing through the coil, can be substantially enhanced in a local region. 
However, this non-linear coil design did not consider the influence of other factors, such as force and torque balance, which are important for a local breast coil design~\cite{lopez_improved_2009}. 

In addition to a high gradient strength, a higher slew rate than a typical clinical system (200T/m/s) is also required to entirely translate the high available gradient strength into relevant clinical applications~\cite{PalmLaun2018}. Minimizing coil inductance or magnetic energy in a coil design helps to increase the maximum available slew rate~\cite{turner_gradient_1993, lopez_improved_2009}. For example, a local gradient coil close to a human-breast has been designed to possess an inductance of less than 100 $\mu$H~\cite{lopez_improved_2009}. This low coil inductance due to small dimensions of the designed coil~\cite{hidalgo-tobon_theory_2010} indicates that very high slew rates as in~\cite{PalmLaun2018} can be easily achieved with existing commercial gradient amplifiers, e.g. with a current of 650 A and a voltage of 1200 V. 

For the case of small-sized gradient coils, a typical design criterion is to minimize power dissipation~\cite{while_theoretical_2013} since high dissipated power can lead to a temperature rise. High temperatures of the coil surface may result in discomfort to the subjects (and ultimately present a risk factor) and therefore are expected to limit the maximum performance of the coil~\cite{chu_mr_1995}. 

In this work, two novel optimization problems are formulated to perform non-linear coil designs, where power dissipation is minimized and the distribution of the resulting encoding field is controlled over the target region and at the same time the potential challenges to the data acquisition and and image reconstruction are taken into account. Force and torque balance, and fabrication constraints, are also considered in the optimization. Due to the unconventional problem formulation without an explicit target field, we propose to convert the multi-objective optimization problem into a constrained optimisation problem, which allows for flexible and direct control of the free model parameters. Three performance parameters are used to assess and compare resulting coil designs. The proposed optimization problems and the performance measures are tested by designing a unilateral single-channel breast gradient coil for diffusion weighting.

\section{Methods}
\label{sec:Methods} 
\subsection{Formulation of the optimizaiton problem}
\begin{figure}[htbp]
	\centering \resizebox{0.9\textwidth}{!}{\includegraphics{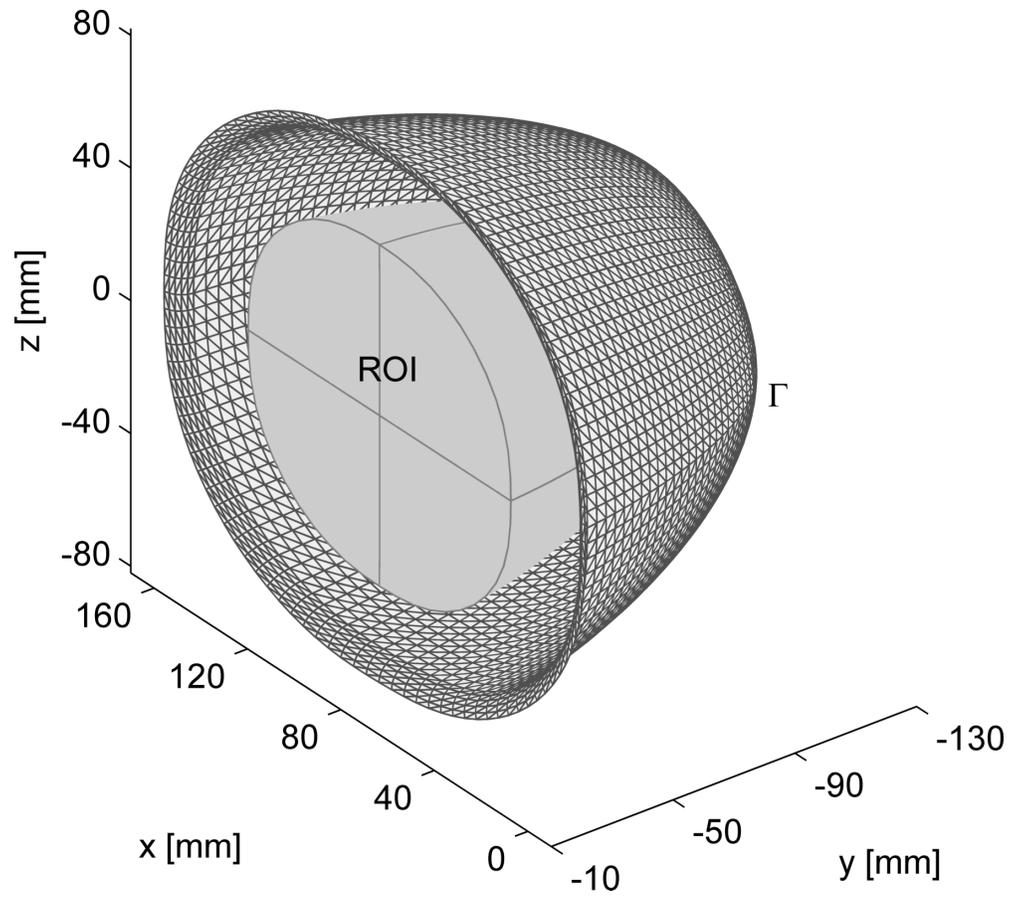}} 
	\caption{Geometries of the ROI and current-carrying surface $\Gamma$. The detailed parameters are presented in Table \ref{table:NBCD_LinearCoils}.}
	\label{fig:CCSMeshROI_fig1} 
\end{figure}

In order to formulate an optimization problem of a non-linear human-breast coil design for diffusion weighting, we first consider a linear coil design. To comply with the requirements of a high gradient strength and low dissipated power, one approach is to maximize the resistive figure of merit (FoM)~\cite{poole_minimax_2012} which is expressed as the gradient strength at the center of a region of interest (ROI) divided by the square root of the power dissipated by the coil. This can be implemented by solving the following problem~\cite{while_theoretical_2013,poole_convex_2014, sanchez_lopez_planar_2018}:
\begin{linenomath}
	\begin{align}
	\min_{\psi}\mbox{ }& \sqrt{P(\psi)},  \mbox{ }P(\psi):=\frac{1}{\tau \sigma}\int_{\Gamma}|\vec{J}(\psi)|^2 d\Gamma ,\label{equ:NBCD_LinearP}\\
	\mbox{subject to } &\mbox{  }
	\max_{\vec{x}_i\in{\text{ROI}}}\left| B_z(\psi,\vec{x}_i)-B_z^*(\vec{x}_i)\right|\leq \delta\max_{\vec{x}_i\in{\text{ROI}}}\left| B_z^*(\vec{x}_i)\right|,\nonumber \\
	&\mbox{ } |F_x|:=\left|\int_{\Gamma}B_0J_y(\psi)d\Gamma\right|\leq F_{\max},\nonumber \\
	&\mbox{ } |F_y|:=\left|-\int_{\Gamma}B_0J_x(\psi)d\Gamma\right|\leq F_{\max},\nonumber \\
	&\mbox{ } |M_x|:=\left|\int_{\Gamma}B_0J_x(\psi)zd\Gamma\right|\leq M_{\max},\nonumber \\
	&\mbox{ } |M_y|:=\left|\int_{\Gamma}B_0J_y(\psi)zd\Gamma\right|\leq M_{\max},\nonumber \\
	&\mbox{ } |M_z|:=\left|-\int_{\Gamma}B_0\left(J_x(\psi)x+J_y(\psi)y\right)d\Gamma\right|\leq M_{\max},\nonumber
	\end{align}
\end{linenomath}
where $\psi$ denotes the scalar piecewise-linear stream function~\citep{peeren_stream_2003} of the electric current density vector $\vec{J}(\psi):=(J_x(\psi),J_y(\psi),J_z(\psi))^T$ and $\vec{J}(\psi) = \nabla\times(\psi\vec{n})$ on a current-carrying surface $\Gamma$ (Fig. \ref{fig:CCSMeshROI_fig1}) with a normal unit vector $\vec{n}$. $P(\psi)$ is power dissipated by the coil, where $\tau$ and $\sigma$ indicate the thickness and the electrical conductivity of the surface $\Gamma$, respectively. The $B_z$ is the z-component of the magnetic field $\vec{B}$ generated by current density $\vec{J}$ on $\Gamma$, which is calculated using the Biot-Savart law. The points $\vec{x}_i = (x_i, y_i, z_i)^T$, $i=1,\ldots, k$, denote the coordinate vectors of $k$ test points in the ROI. The $B_z^*$ is a given linear target magnetic field. $\delta$ is a given constant to control the linearity of the resulting $B_z$, which is specified as to be 5 \% as usual~\cite{poole_convex_2014}. 

Due to the missing symmetry of the surface $\Gamma$ with respect to the direction of $B_0$ and isoisocentre of the magnet, it is important to control forces and unbalanced torques on the designed coil. In Problem (\ref{equ:NBCD_LinearP}), $F_x$ and $F_y$ indicate x- and y-components of the net force on the coil, respectively, in the presence of a main magnetic field $B_0$ along z-direction. In the examples of this manuscript, the main field $B_0$ is assumed to be homogenous with the strength of 3 T. Therefore, the constraints of $F_x$ and $F_y$ are naturally satisfied and can be ignored during the optimization of Problem (\ref{equ:NBCD_LinearP}). $M_x$, $M_y$ and $M_z$ denote x-, y- and z-components of the magnetic torque executed on the coil, respectively. $F_{\max}$ and $M_{\max}$ are two given positive constants with the unit of N and N$\cdot$m in order to set a limitation of the total force and the total magnetic torque, respectively.

For non-linear coil designs, a corresponding resistive figure of merit can also be used to balance the tradeoff between increasing gradient strength and reducing power dissipation. Unlike a linear coil design, the resulting magnetic field from a non-linear coil is not required to have a linear shape. Instead of the requirement of linearity, an encoding field generated by a designed non-linear coil should have strong local gradients at every test point within an ROI. 

For non-linear coils, gradient strength of the encoding fields typically varies spatially within the ROI. In order to obtain a useful non-linear encoding field within the ROI, it makes sense to sum the gradient strength of the encoding field over all the test points in the ROI and consider the sum in the definition of a resistive figure of merit for a non-linear coil. One definition of the resistive FoM, with the unit of T/(m$\cdot$A$\cdot\Omega^{1/2}$), was introduced in~\cite{jia_design_2017} as follows:
\begin{linenomath}
	\begin{align}
	\beta_P&:=\frac{\sum_{\vec{x}_i\in{\text{ROI}}}\lvert\nabla B_z(\psi,\vec{x}_i)\rvert}{k\sqrt{P(\psi)}}. \label{equ:NBCD_beta}  
	\end{align}
\end{linenomath}  
Here, larger $\beta_P$ of a coil implies higher average gradient strength over the ROI and a better performance for a given power dissipated by the coil. Note that Equation (\ref{equ:NBCD_beta}) can also be used for calculations of the resistive FoM for a linear coil in the latter examples to allow for fair comparisons between the performance of linear and non-linear coils.

In order to maximize the resistive figure of merit (\ref{equ:NBCD_beta}) for a non-linear coil, the following optimization problem is proposed based on a hint from Problem (\ref{equ:NBCD_LinearP}):
\begin{linenomath}
	\begin{align}
	\min_{\psi}\mbox{ }& F,  \mbox{ }F:=\sqrt{P(\psi)}+ \alpha_J \lVert\vec{J}(\psi)\rVert_{p} ,\label{equ:NBCD_non-linearPJ}\\
	\mbox{subject to } &\mbox{  } \frac{1}{k}\sum_{\vec{x}_i\in{\text{ROI}}}\left|\nabla B_z(\psi,\vec{x}_i)\right|\geq C_g, \nonumber\\
	&\mbox{ } |F_x|:=\left|\int_{\Gamma}B_0J_y(\psi)d\Gamma\right|\leq F_{\max},\nonumber \\
	&\mbox{ } |F_y|:=\left|-\int_{\Gamma}B_0J_x(\psi)d\Gamma\right|\leq F_{\max},\nonumber \\
	&\mbox{ } |M_x|:=\left|\int_{\Gamma}B_0J_x(\psi)zd\Gamma\right|\leq M_{\max},\nonumber \\
	&\mbox{ } |M_y|:=\left|\int_{\Gamma}B_0J_y(\psi)zd\Gamma\right|\leq M_{\max},\nonumber \\
	&\mbox{ } |M_z|:=\left|-\int_{\Gamma}B_0\left(J_x(\psi)x+J_y(\psi)y\right)d\Gamma\right|\leq M_{\max}.\nonumber
	\end{align}
\end{linenomath}
Here, $\alpha_J$ and $C_g$ are two positive constants. In contrast to Problem (\ref{equ:NBCD_LinearP}), the p-norm of the surface current density $\vec{J}(\psi)$ ($\lVert J(\psi) \rVert_p:=(\int_{\Gamma}|\vec{J}(\psi)|^p d\Gamma)^{1/p} (p>2)$)~\cite{JiaZaitsev2016a} is selected as one of the objectives in Problem (\ref{equ:NBCD_non-linearPJ}) to modify the width of wire tracks for a practical fabrication of the coil. The weight $\alpha_J$ is tuned to meet the requirement of the minimum wire width. The selection of the integer $p$ 
depends on coil design objectives and the scaling of the problem. For the current breast coil design $p=7$ was determined experimentally and used throughout this work. As described in Problem (\ref{equ:NBCD_non-linearPJ}), power dissipation for a non-linear coil is minimized while the average of the gradient strength of the resulting magnetic field $B_z$ over the ROI is constrained by a lower bound $C_g$ in order to maximize the resistive figure of merit $\beta_P$.

Unlike Problem (\ref{equ:NBCD_LinearP}) for a linear coil design, Problem (\ref{equ:NBCD_non-linearPJ}) contains no specification of the shape of the desired magnetic field. Preliminary investigations have shown a tendency of Problem (\ref{equ:NBCD_non-linearPJ}) to favour solutions offering high gradients accompanied by a strong non-linearity of the encoding field. When such fields are used to generate diffusion weighing, due to the quadratic dependency of the b-factor on the local gradient strength, methodological difficulties were anticipated. Indeed, if the local gradients within a single imaging slice differ by a significant factor, it may turn out that in some areas of the slice the induced diffusion weighting is too weak to produce any usable signal attenuation, whereas in others it it too strong and will completely suppress MR signals. A possible solution to use more diffusion weighting steps in a multi-b-factor protocol. While being feasible, this may result in a substantial extension of the imaging times. Therefore a decision was made to optimize the above-mentioned aspects during the coil design by adding further constraints to the optimisation problem. We propose to require the encoding field generated by a non-linear breast coil to have a nearly constant gradient magnitude, independent of the direction, within each coronal slice.  Therefore we allow only for such field non-linearities, that lead to variation of the gradient strength along y axis or to varying gradient direction. This requirement will eventually allow us to employ the new coil for diffusion weighting within each of the coronal slices in a way that is similar to the currently established protocols that rely in linear gradients, possibly leading to faster acquisition and easier and more robust image reconstruction and data processing.
Based on the above requirements, another optimization problem is proposed to design a non-linear breast coil as follows:
\begin{linenomath}
	\begin{align}
	\min_{\psi}\mbox{ }& F,  \mbox{ }F:=\sqrt{P(\psi)}+ \alpha_J \lVert\vec{J}(\psi)\rVert_{p} ,\label{equ:NBCD_non-linearF}\\
	\mbox{subject to } &\mbox{  } \frac{1}{k}\sum_{\vec{x}_i\in{\text{ROI}}}\left|\nabla B_z(\psi,\vec{x}_i)\right|\geq C_g, \nonumber \\
	&\mbox{  }\frac{1}{k\sqrt{P(\psi)}}\left( \sum_{\vec{x}_i\in{\text{ROI}}}\left(\frac{\left|\nabla B_z(\psi,\vec{x}_i)\right|}{\partial x}\right)^2+\left(\frac{\left|\nabla B_z(\psi,\vec{x}_i)\right|}{\partial z}\right)^2 \right)^\frac{1}{2}\leq C_c, \label{equ:NBCD_CoronalSliceConstraints}\\
	&\mbox{ } |F_x|:=\left|\int_{\Gamma}B_0J_y(\psi)d\Gamma\right|\leq F_{\max},\nonumber \\
	&\mbox{ } |F_y|:=\left|-\int_{\Gamma}B_0J_x(\psi)d\Gamma\right|\leq F_{\max},\nonumber \\
	&\mbox{ } |M_x|:=\left|\int_{\Gamma}B_0J_x(\psi)zd\Gamma\right|\leq M_{\max},\nonumber \\
	&\mbox{ } |M_y|:=\left|\int_{\Gamma}B_0J_y(\psi)zd\Gamma\right|\leq M_{\max},\nonumber \\
	&\mbox{ } |M_z|:=\left|-\int_{\Gamma}B_0\left(J_x(\psi)x+J_y(\psi)y\right)d\Gamma\right|\leq M_{\max}.\nonumber
	\end{align}
\end{linenomath}
Here, $C_c$ is a positive constant. The constraint (\ref{equ:NBCD_CoronalSliceConstraints}) is used to homogenize the gradient strength of $B_z$ on each coronal slices.

To guarantee that resulting wire patterns are closed on the current-carrying surface $\Gamma$, all values of the stream function $\psi$ must remain constant on each closed boundary of $\Gamma$ ~\citep{peeren_stream_2003}. Moreover, $\psi$ is typically specified as a given constant, such as 0, at one point in $\Gamma$ or on one closed boundary of $\Gamma$ to avoid the existence of infinitely many solutions of Problems (\ref{equ:NBCD_LinearP}) and (\ref{equ:NBCD_non-linearF}). Without this requirement for any stream function $\psi$, which is a solution of Problem (\ref{equ:NBCD_LinearP}), $\psi$ plus any constant is also a solution of Problem (\ref{equ:NBCD_LinearP}). As shown in Figure \ref{fig:CCSMeshROI_fig1}, the current-carrying surface $\Gamma$ has only one boundary and $\psi$ on this boundary is set to 0 in all subsequent numerical examples.

\subsection{Proposed performance measure}
In addition to the resistive figure of merit $\beta_P$, two extra performance metrics are proposed to assess and compare different breast coil designs. The first metric $\nu_P$ is introduced to compare the inhomogeneity of the gradient amplitude of $B_z$ in coronal slices for a given dissipated power. Secondly, an inductive figure of merit $\beta_W$ is intended to measure the coil performance from the point of view of the stored magnetic energy, which is directly related to the switching rate. Based on the above considerations, the formulae of $\nu_P$ and $\beta_W$ are defined as follows:  
\begin{linenomath}
	\begin{align}
	\nu_P&:=\frac{1}{k\sqrt{P(\psi)}}\left( \sum_{\vec{x}_i\in{\text{ROI}}}\left(\frac{\left|\nabla B_z(\psi,\vec{x}_i)\right|}{\partial x}\right)^2+\left(\frac{\left|\nabla B_z(\psi,\vec{x}_i)\right|}{\partial z}\right)^2 \right)^\frac{1}{2}, \label{equ:NECD_nuP} \\
	\beta_W&:=\frac{\sum_{\vec{x}_i\in{\text{ROI}}}\lvert\nabla B_z(\psi,\vec{x}_i)\rvert}{k\sqrt{W_m}}. \label{equ:NBCD_betaW}  
	\end{align}
\end{linenomath}                                         
Here, $W_m:=\frac{\mu_0}{8\pi}\int_{\Gamma}\int_{\Gamma'}\frac{\vec{J}(\psi(\vec{x}))\cdot\vec{J}(\psi(\vec{x}'))}{|\vec{x}-\vec{x}'|} d\Gamma'd\Gamma$~\cite{lemdiasov_stream_2005} indicates coil magnetic energy and $\mu_0$ denotes the magnetic constant of $4\pi\times10^{-7}$ H/m. 

As shown in Equation (\ref{equ:NECD_nuP}), $\nu_P$ with the unit of T/(m$^2\cdot$A$\cdot\Omega^{1/2}$) is defined as the left term of Constraint (\ref{equ:NBCD_CoronalSliceConstraints}). For given dissipated power, lower $\nu_P$ indicates lower average variation of the gradient strength within coronal slices, leading to an easier data acquisition and image reconstruction in coronal slice orientations. As presented in Equation (\ref{equ:NBCD_betaW}),  $\beta_W$ with the unit of T/(m$\cdot$A$\cdot$H$^{1/2}$) is described as the average of gradient strength of the encoding field over the ROI divided by the square root of the magnetic energy stored by the coil.  Thus, for given magnetic energy, larger $\beta_W$ of a coil implies higher average gradient strength and a better performance of the coil.

\subsection{Numerical optimization procedure and parameters}
In order to obtain a non-linear breast coil comparable to its corresponding linear coil, the whole optimization procedure is divided into three steps. In the first step, a linear breast coil is designed using Problem (\ref{equ:NBCD_LinearP}) when the target field $B_z^*$ is provided. Considering that Problem (\ref{equ:NBCD_LinearP}) is equivalent to a quadratic optimization problem where the objective is specified as $P_d$ rather than $\sqrt{P_d}$ and the constraints remain the same, the solution of Problem (\ref{equ:NBCD_LinearP}) is obtained by solving the equivalent problem with the function $quadprog$ from MATLAB (The MathWorks. Natick, USA). Secondly, Problems (\ref{equ:NBCD_non-linearPJ}) and (\ref{equ:NBCD_non-linearF}) with $\alpha_J$ = 0 and using the stream function resulting from the first step as an initial value, are solved to obtain a non-linear breast coil, respectively. Finally, realizable non-linear coils are obtained by solving Problems (\ref{equ:NBCD_non-linearPJ}) and (\ref{equ:NBCD_non-linearF}) with $\alpha_J$s tuned to satisfy with the requirement of minimum wire widths of the coils. The stream function resulting from the second step is used as the initial value at this step. Moreover, the left formulae of the first two constraints of Problem (\ref{equ:NBCD_non-linearF}) are calculated using the resulting magnetic field $B_z$ generated by the designed linear coil. Then, $C_g$ and $C_c$ of Problem (\ref{equ:NBCD_non-linearF}) are set to be equal to these two calculated values, respectively. Finally, realizable non-linear coils are obtained by solving Problem (\ref{equ:NBCD_non-linearPJ}) and (\ref{equ:NBCD_non-linearF}) with tuned $\alpha_J$s to satisfy with the requirement of minimum wire widths of the coils. The initial value at this step is the resulting stream function from the second step and $C_g$ and $C_c$ remain the same as those in the second step. All numerical examples for non-linear coil designs were solved using the function $fmincon$ from MATLAB (The MathWorks. Natick, USA).

In these examples, the current-carrying surface $\Gamma$ (Fig.~\ref{fig:CCSMeshROI_fig1}) has the shape of a cup. The bottom parts of the surface $\Gamma$ and the boundary of the ROI along y-axis have a similar superellipical shape. The ROI is defined as the intersection of the closed three-dimentional half-space \{$(x, y, z)|y\leq-25 \mbox{ mm}$.\} and the superellipsoid defined by \{$(x, y, z)|(|(x-x_0)/A_s|^r+|y/B_s|^r)^{t/r}+|z/C_s|^t\leq 1.$\}, where $x_0$, $A_s$, $B_z$, $C_s$, $r$ and $t$ are equal to 83 mm, 57 mm, 102 mm, 52 mm, 2.5 and 2.5, respectively. The bottom part of $\Gamma$ has been obtained by upscaling the bottom of the ROI boundary. The distance between a point on the ROI boundary and its corresponding point on $\Gamma$ is specified as 24 mm to reserve sufficient space to insert an RF coil between the gradient coil and the sample. The volume of the ROI 
is chosen to be large enough to accomodate a large fraction of the female population.
 The thickness, $\tau$, and conductivity, $\sigma$, are set as 6 mm and  $5.998\times10^4$ S/mm, respectively. $F_{\max}$ and $M_{\max}$ are specified as $1\times10^{-4}$ N and $1\times10^{-4}$ N$\cdot$m, respectively, to limit the total net magnetic force and torque of a designed coil.

\section{Results}
\label{sec:Results} 
This section includes three subsections. In the first subsection, we designed three linear gradient coils on the original current-carrying surface $\Gamma$ (Fig. \ref{fig:CCSMeshROI_fig1}) by solving Problem (\ref{equ:NBCD_LinearP}) as detailed below. Based on the resulting designs, we selected one solution on $\Gamma$ as an initial value of Problem (\ref{equ:NBCD_non-linearPJ}) and (\ref{equ:NBCD_non-linearF}). Secondly, we solved Problem (\ref{equ:NBCD_non-linearPJ}) and (\ref{equ:NBCD_non-linearF}) to obtain two non-linear gradient coils without and with the constraint of the resulting gradient field within coronal slices, respectively. Finally, another extended current-carrying surface was used to design non-linear coils in order to improve performance of the optimal coils obtained in the second subsection. 


\subsection{Linear gradient coil design}
\label{subsec:LinearCoils}
Table \ref{table:NBCD_LinearCoils} shows a performance comparison of three linear gradient coils designed by solving Problem \ref{equ:NBCD_LinearP} using linear $G_x$, $G_y$ and $G_z$ target fields with the same gradient strength of 50 mT/m. This comparison indicates that the linear $G_z$ coil outperforms linear $G_x$ and $G_y$ coils and can be used as a suitable initial value of subsequent non-linear coil designs. As can be seen, the required currents flowing through the designed $G_x$ and $G_y$ coils have been raised 2.29 and 43.7 times, respectively, as compared with the designed $G_z$ coil. These increases are linked to the corresponding decreases in coil efficiency $\eta$ for the $G_x$ and $G_y$ coils. Here, $\eta$ is defined as the ratio of the magnitude of the gradient of the encoding field $B_z$ to the current flowing through the coil at sampling points in an ROI. In this way, the concept of $\eta$ can also be used in later non-linear coils. Moreover, compared with the $G_z$ coil, the resistive figure of merit $\beta_P$ for the $G_x$ and $G_y$ coils drop 5.55 and 238 fold, respectively, while the corresponding inductive figures of merit $\beta_W$ decreases 2.17 and 77.5 times, respectively. One may notice relatively strong variation of the local gradient efficiency (the ratio between the local gradient strength and the applied current), especially for $G_x$ (about factor of 2.55) and $G_z$ (about factor of 2.15). This is due to the unusual non-symmetric shapes of both the current carrying surface and the target region.


\subsection{Non-linear gradient coil without and with constraints within coronal slices}
\label{subsec:NonLinearCoilsOnGamma}
Table \ref{table:NBCD_Non-LinearCoils} displays properties of two non-linear coil designs on the same current-carrying surface $\Gamma$. One was obtained by solving Problem (\ref{equ:NBCD_non-linearF}) without Constraint (\ref{equ:NBCD_CoronalSliceConstraints}) within coronal slices and the other is the solution of Problem (\ref{equ:NBCD_non-linearPJ}) with Constraint (\ref{equ:NBCD_CoronalSliceConstraints}). As listed, the same limiting constants $C_g$ was used and different $\alpha_J$s are selected to achieve the same minimum wire width of 2.5 mm in both cases. Here, the optimal stream function for the linear $G_z$ coil design in the above subsection \ref{subsec:LinearCoils} was specified as the initial value of these two non-linear cases. Constraints $C_g$ and $C_c$ were calculated by using the encoding field generated by the linear $G_z$ coil.

Table \ref{table:NBCD_Non-LinearCoils} also presents a performance comparison of these two optimized non-linear coils. As shown, the non-linear coil designed with Constraint (\ref{equ:NBCD_CoronalSliceConstraints}) has its own features. For example, although the resistive and inductive figures of merit for the non-linear coil with Constraint (\ref{equ:NBCD_CoronalSliceConstraints}) decrease by around 1.49 and 1.40 times, respectively, as compared to the coil without this constraint, the inhomogeneity metric $\nu_P$ of the gradient strength within coronal slices reduces by a factor of 9.80, thus leading to the same value of $\nu_P$ as for the linear $G_z$ coil with 5\% field error (Table \ref{table:NBCD_LinearCoils}). 

Additionally, the non-linear coil with Constraint (\ref{equ:NBCD_CoronalSliceConstraints}) also outperforms the linear $G_z$ coil in several ways, as presented in Tables \ref{table:NBCD_LinearCoils} and \ref{table:NBCD_Non-LinearCoils}. First, compared with the linear coil, the resistive figure of merit $\beta_P$ for the non-linear coil raises 4.14 times. This increase is due to the reduction of dissipated power since the integrated gradient
strength of the resulting encoding fields does not vary for these two coils. Next, the inductive figure of merit $\beta_W$ for the non-linear coil increases 3.98-fold although $\beta_W$ is not directly optimized in the proposed design method. Finally, the non-linear coil had a 7.40 times as high maximum local coil efficiency $\eta$ (local gradient strength per unit current) as the linear coil.

\begin{figure}[htbp]
	\centering \subfigure[]{ \label{fig:EtaXfemNoBrim_fig2a} \includegraphics[width=0.425\textwidth]{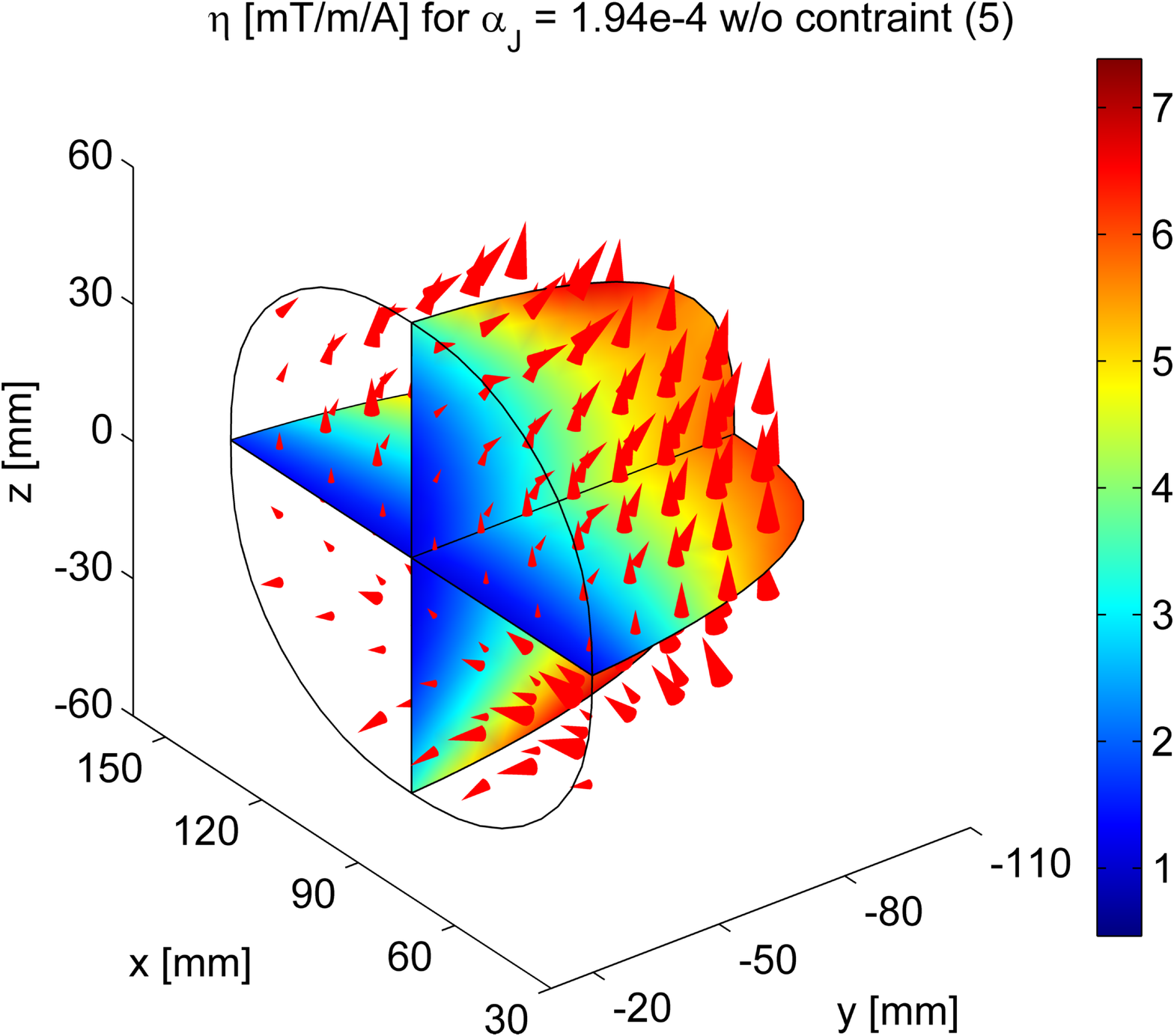}
	} \subfigure[]{ \label{fig:EtaXfemNoBrim_fig2b} \includegraphics[width=0.425\textwidth]{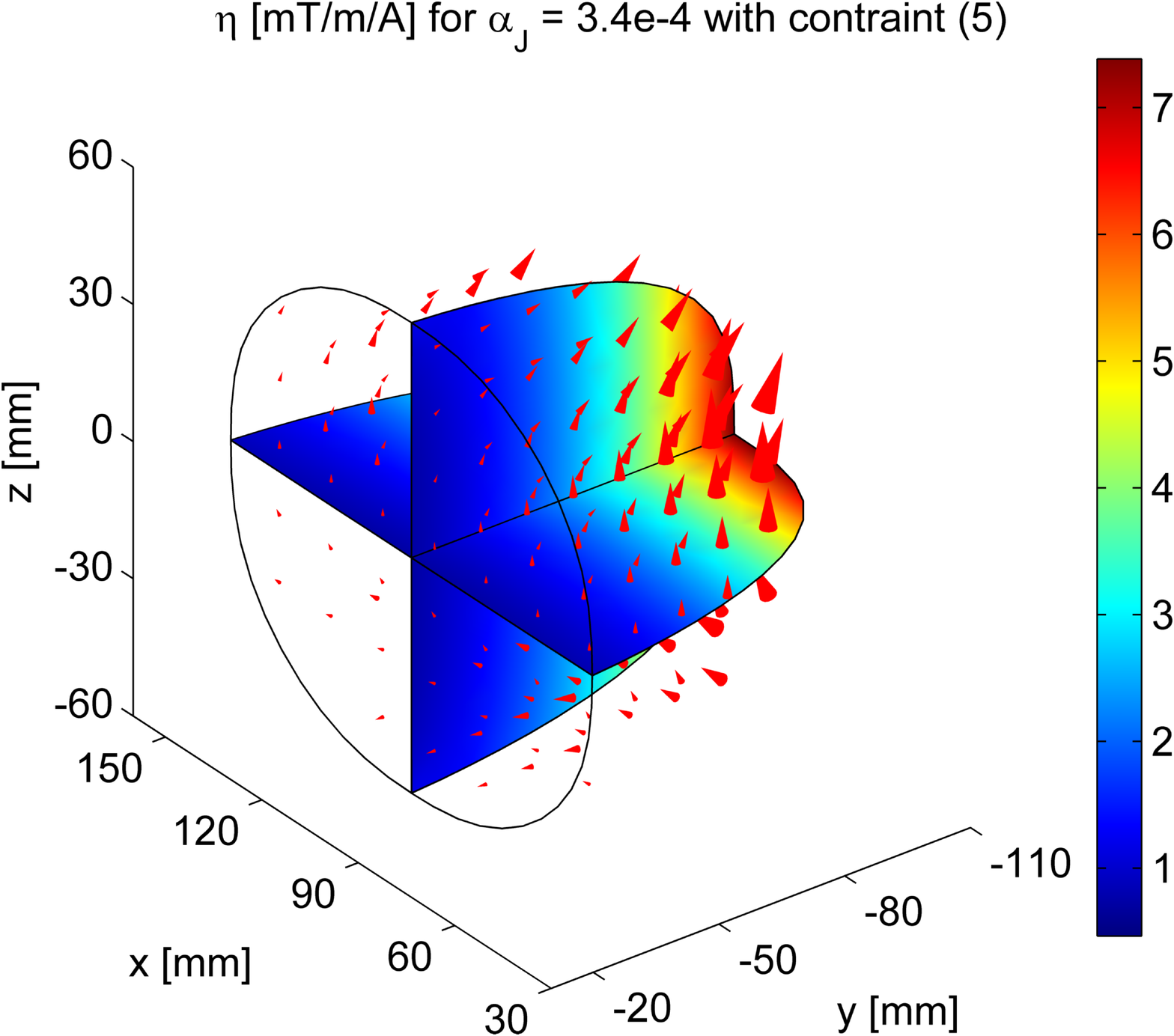}
	} \subfigure[]{ \label{fig:EtaXfemNoBrimLinear_fig2c} \includegraphics[width=0.425\textwidth]{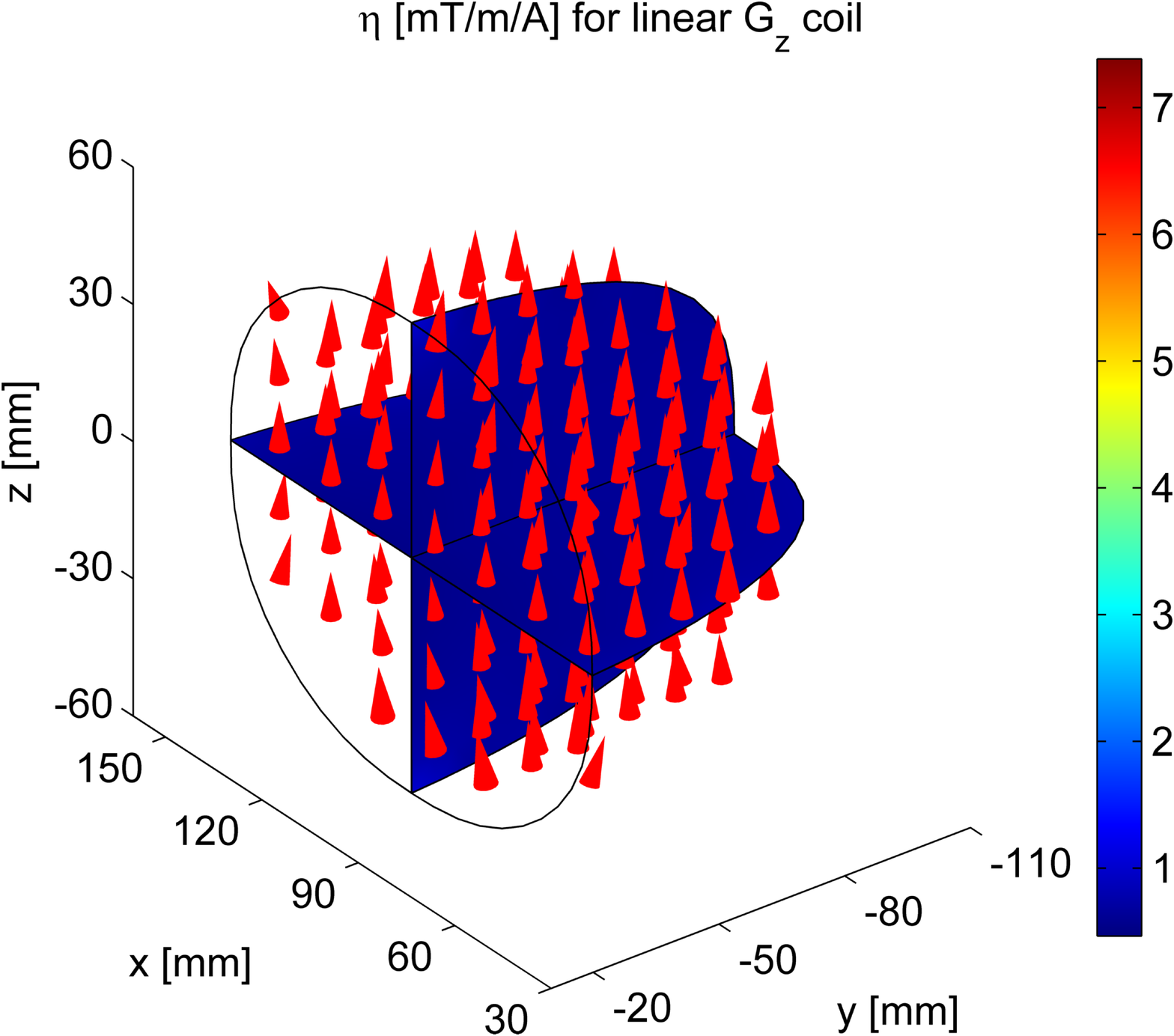}
	} \subfigure[]{ \label{fig:HistCompROINoBrim_fig2d} \includegraphics[width=0.425\textwidth]{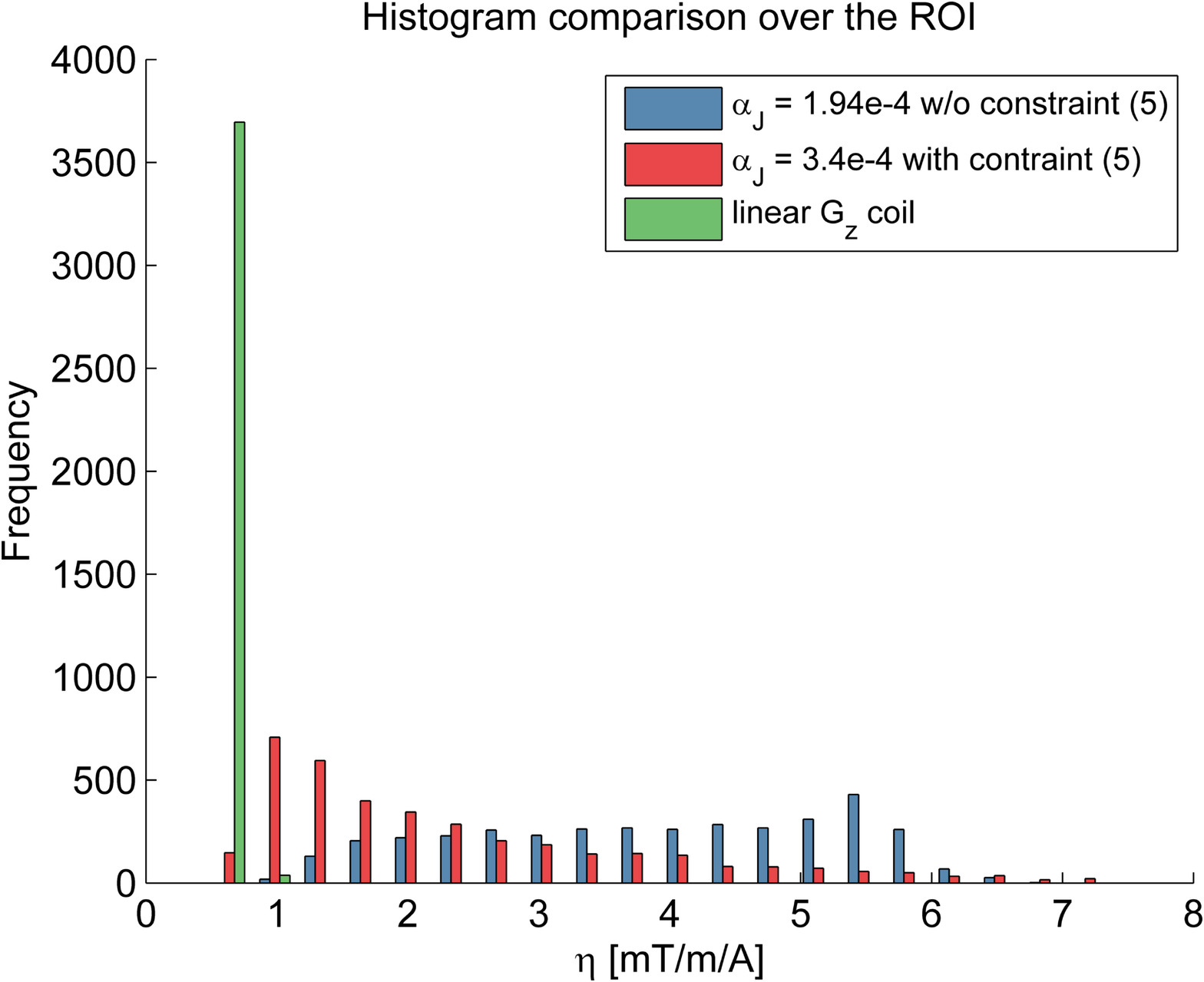}
	} \subfigure[]{ \label{fig:HistCompS1NoBrim_fig2e} \includegraphics[width=0.425\textwidth]{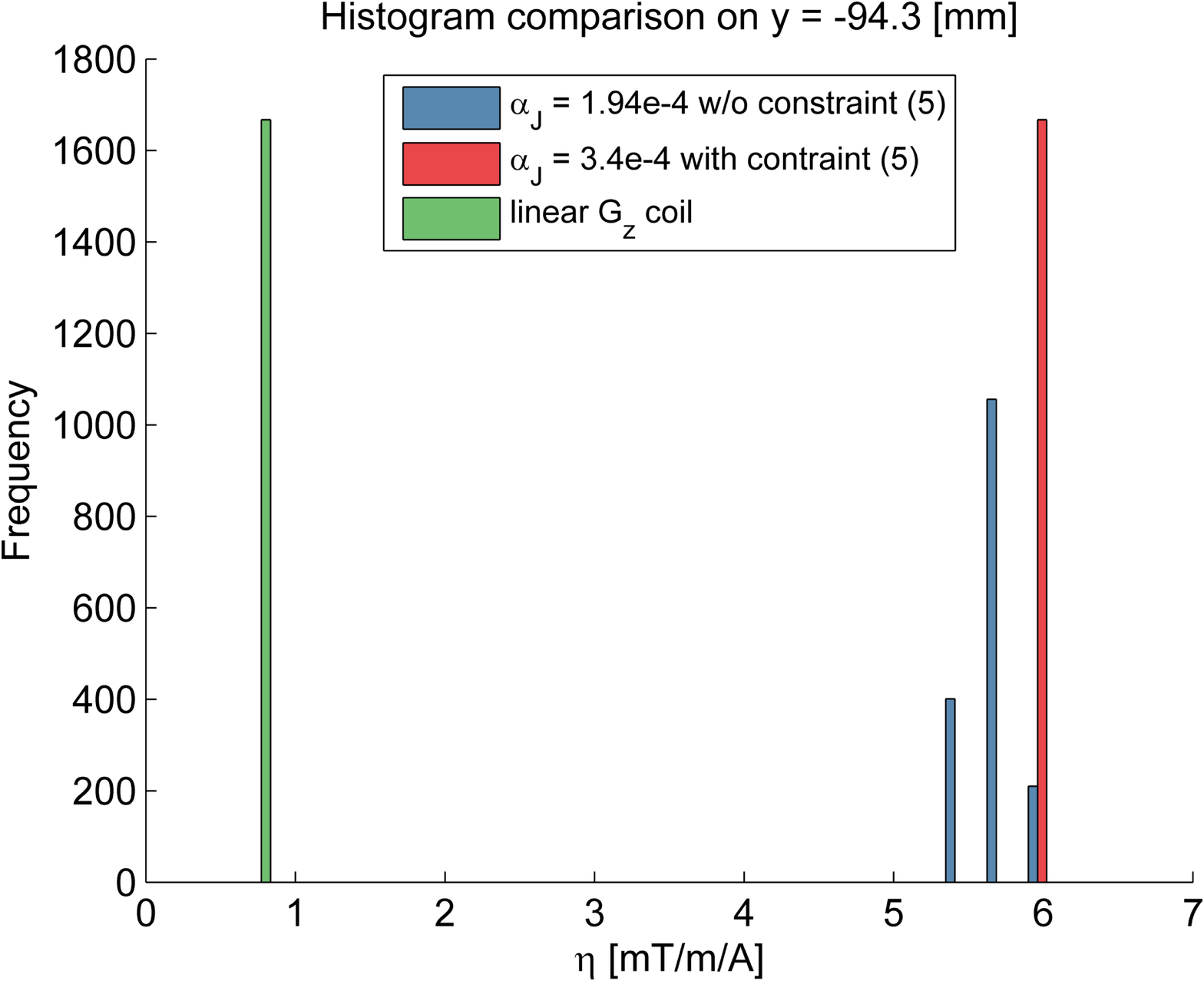}
	} \subfigure[]{ \label{fig:HistCompS5NoBrim_fig2f} \includegraphics[width=0.425\textwidth]{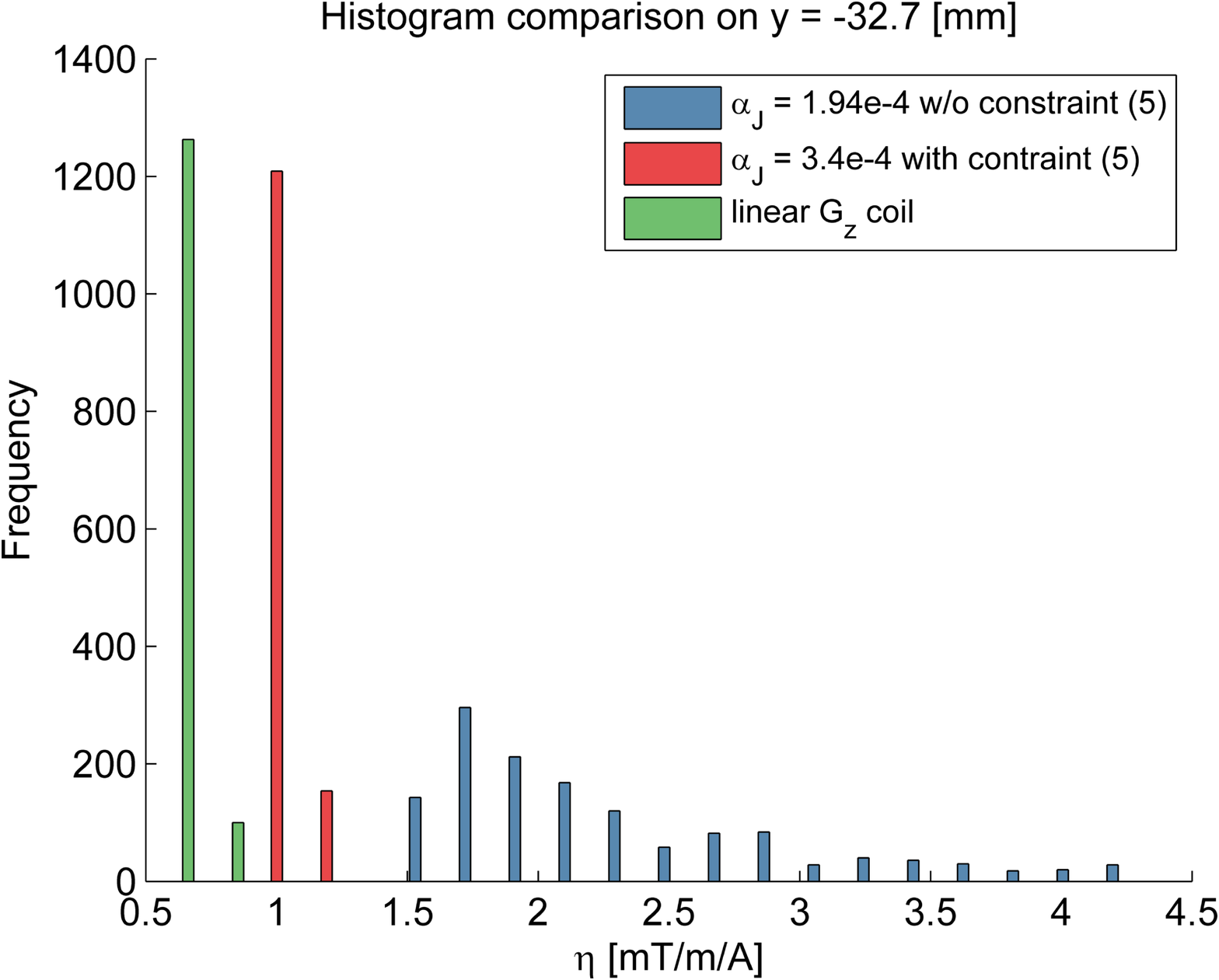}
	}\caption{Coil efficiency $\eta$ for the designed non-linear coils without (a) and with (b) constraint ($\ref{equ:NBCD_CoronalSliceConstraints}$). In order to reveal high performance of the non-linear coils, coil efficiency $\eta$ for the corresponding linear $G_z$ coil (c) is also plotted. Here, Here, gradient vectors of the $B_z$ field generated by the three coils with one unit current are marked in red arrows and the arrow lengths are proportional to the magnitude of the gradient vectors. The histogram (d) presents a rough probability distribution of coil efficiency $\eta$ over the ROI for these three coils. The histograms (e) and (f) show the distribution of $\eta$ within two representative coronal slices of $y=-94.3$ mm and $y=-32.7$ mm, respectively.}
	\label{fig:EtaXfemNoBrim_fig2} 
\end{figure}

Figure \ref{fig:EtaXfemNoBrim_fig2} shows the distribution of coil efficiency $\eta$ over the ROI and on two representative coronal slices for the two non-linear coils and the linear $G_z$ coil. As can be seen in Figure \ref{fig:EtaXfemNoBrim_fig2}a-c, the coil efficiency $\eta$ for these two non-linear coils has a higher value at every sampling point in the ROI than that for the linear coil although the minimum of $\eta$ for the non-linear coil given by Constraint (\ref{equ:NBCD_CoronalSliceConstraints}) is lower than the maximum of $\eta$ for the linear coil (Figure \ref{fig:EtaXfemNoBrim_fig2}.d). Actually, compared to the linear coil, coil efficiencies for the non-linear coils without and with Constraint (\ref{equ:NBCD_CoronalSliceConstraints}) increase by at least 1.70 and 1.09 times, respectively. Moreover, in most parts of the ROI (except for the area far away from the chest), the non-linear coil without the constraint outperforms that with the constraint in terms of the coil efficiency. However, as depicted in Figure \ref{fig:EtaXfemNoBrim_fig2}.d, coil efficiencies for these two non-linear coil have higher standard deviations over the ROI than the linear coil. Gradient strength isosurfaces of the non-linear coil without Constraint (\ref{equ:NBCD_CoronalSliceConstraints}) are strongly curved in three dimensions (Figure \ref{fig:EtaXfemNoBrim_fig2}.a), which is expected to pose a challenge to a data acquisition and image reconstruction for DWI. 

Unlike the non-linear coil without Constraint (\ref{equ:NBCD_CoronalSliceConstraints}), $\eta$ for the non-linear coil with the proposed constraint tends to remain constant in every coronal slices (Figure \ref{fig:EtaXfemNoBrim_fig2}.b), possibly helping to overcome the above-mentioned challenges if coronal slice orientation is used. For examples, as illustrated in Figure \ref{fig:EtaXfemNoBrim_fig2}.e and \ref{fig:EtaXfemNoBrim_fig2}.f, $\eta$ for the non-linear coil with Constraint (\ref{equ:NBCD_CoronalSliceConstraints}) tends to have a distribution with similar dispersion as the linear coil within two representative coronal slices at $y=-94.3$ mm and $y=-32.7$ mm, while coil efficiency for the non-linear coil without the constraint has a distribution with a substantially higher standard deviation than other two coils. Actually, the ratios of the standard deviation to the mean of $\eta$  in the coronal slice of $y=-32.7$ mm are 0.093 and 0.066 for the linear coil and the non-linear coil with Constraint (\ref{equ:NBCD_CoronalSliceConstraints}), respectively. However, the ratio in the slice for the non-linear coil without the constraint increases by 3.12 and 4.40 times compared to these two coils, respectively.
Moreover, compared to the linear coil, the mean of $\eta$ for the non-linear coil with the constraint increases more than 9.13-fold in the slice at $y=-94.3$ mm. Particularly, the current of only 170 A and the dissipated power of 516 W are required to achieve the gradient strength of 1T/m in this slice using the non-linear coil. In order to obtain the same gradient strength using the linear coil, more than 9.13 times current and 100-fold dissipated power would be needed, for which the required water-cooling would be infeasible.

\begin{figure}[htbp]
	\centering \subfigure[]{ \label{fig:SFCoilLayout1NoBrim_fig3a} \includegraphics[width=0.31\textwidth]{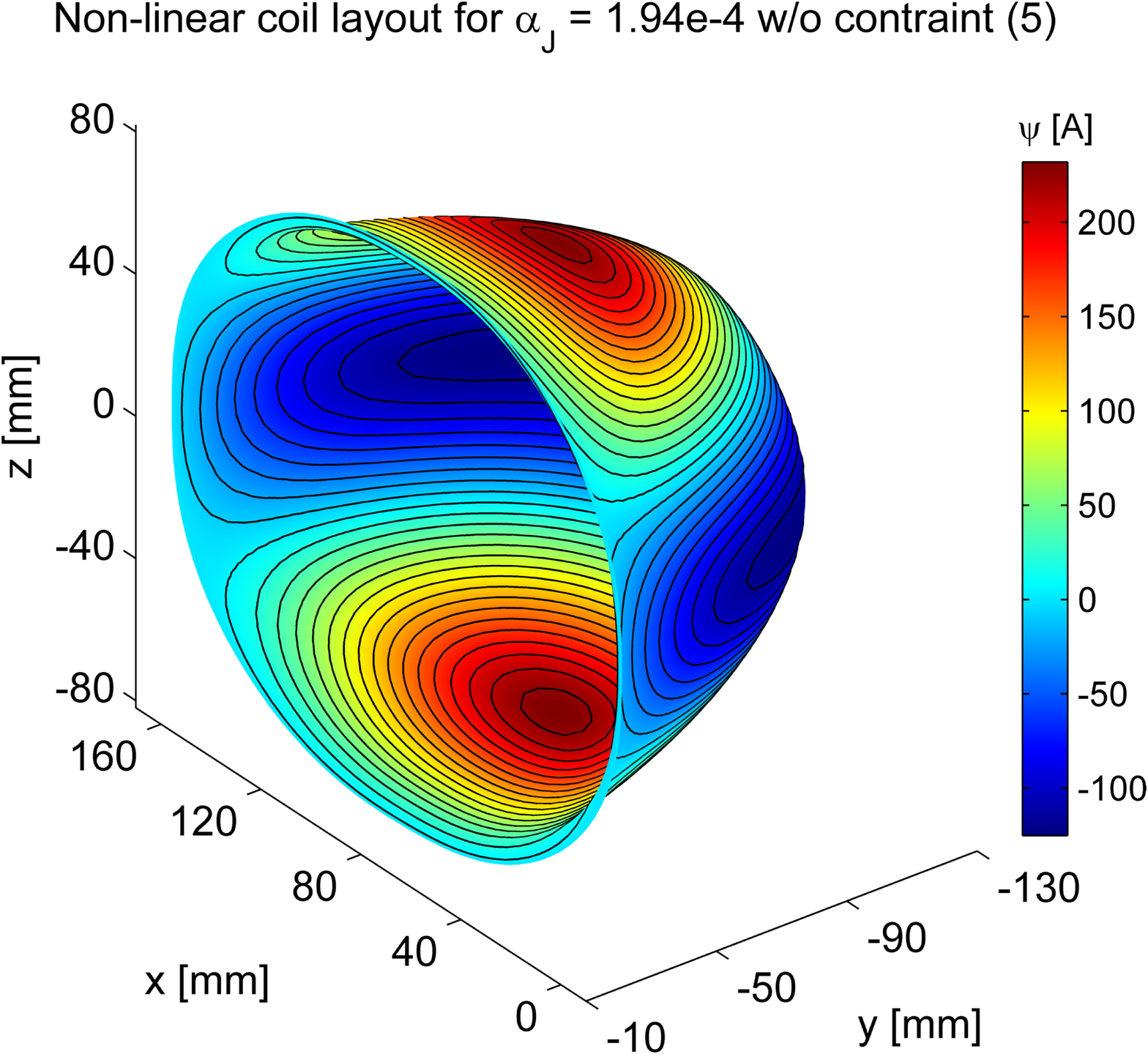}
	} \subfigure[]{ \label{fig:SFCoilLayout2NoBrim_fig3b} \includegraphics[width=0.31\textwidth]{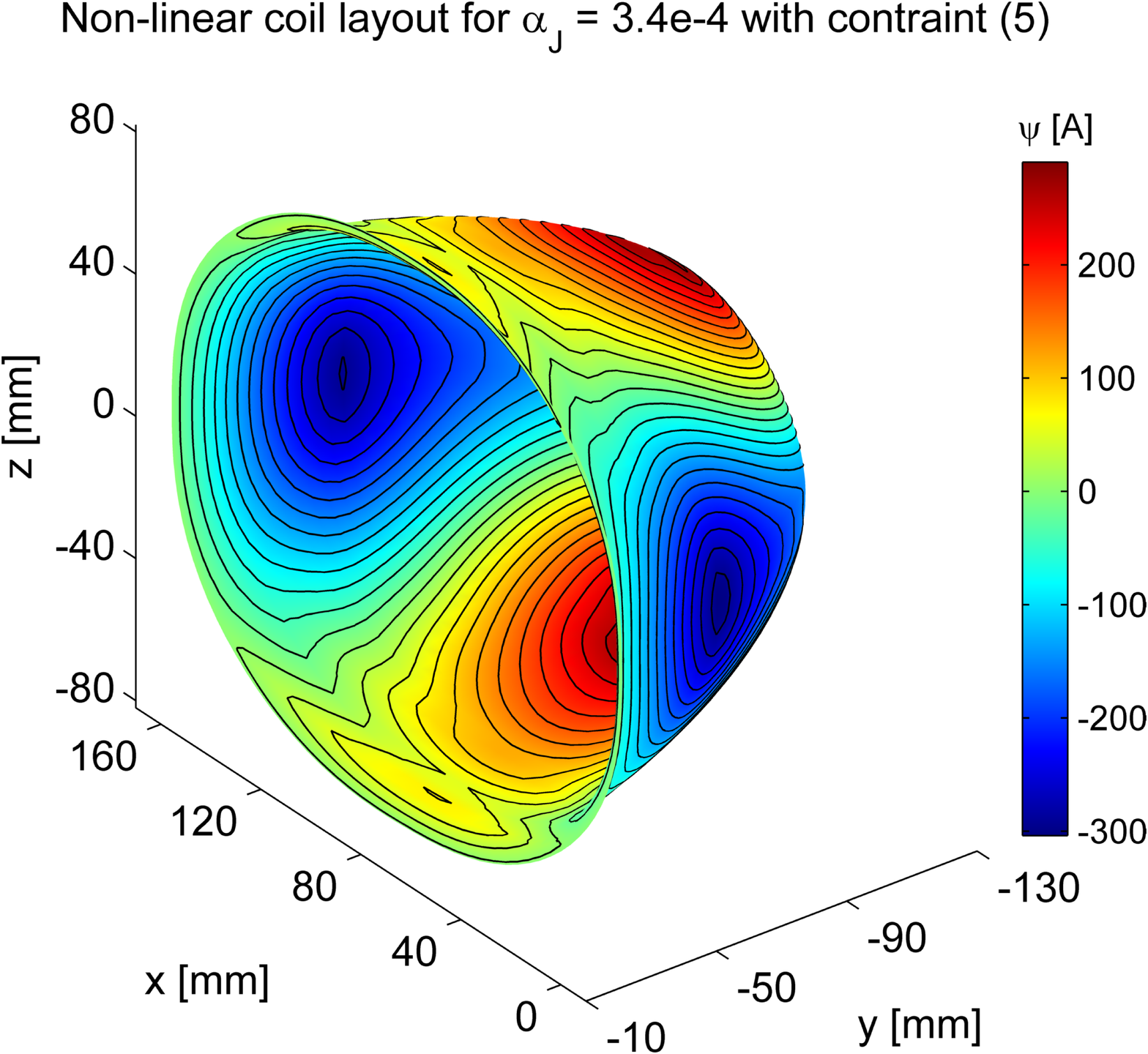}
	} \subfigure[]{ \label{fig:SFCoilLayout3NoBrim_fig3c} \includegraphics[width=0.31\textwidth]{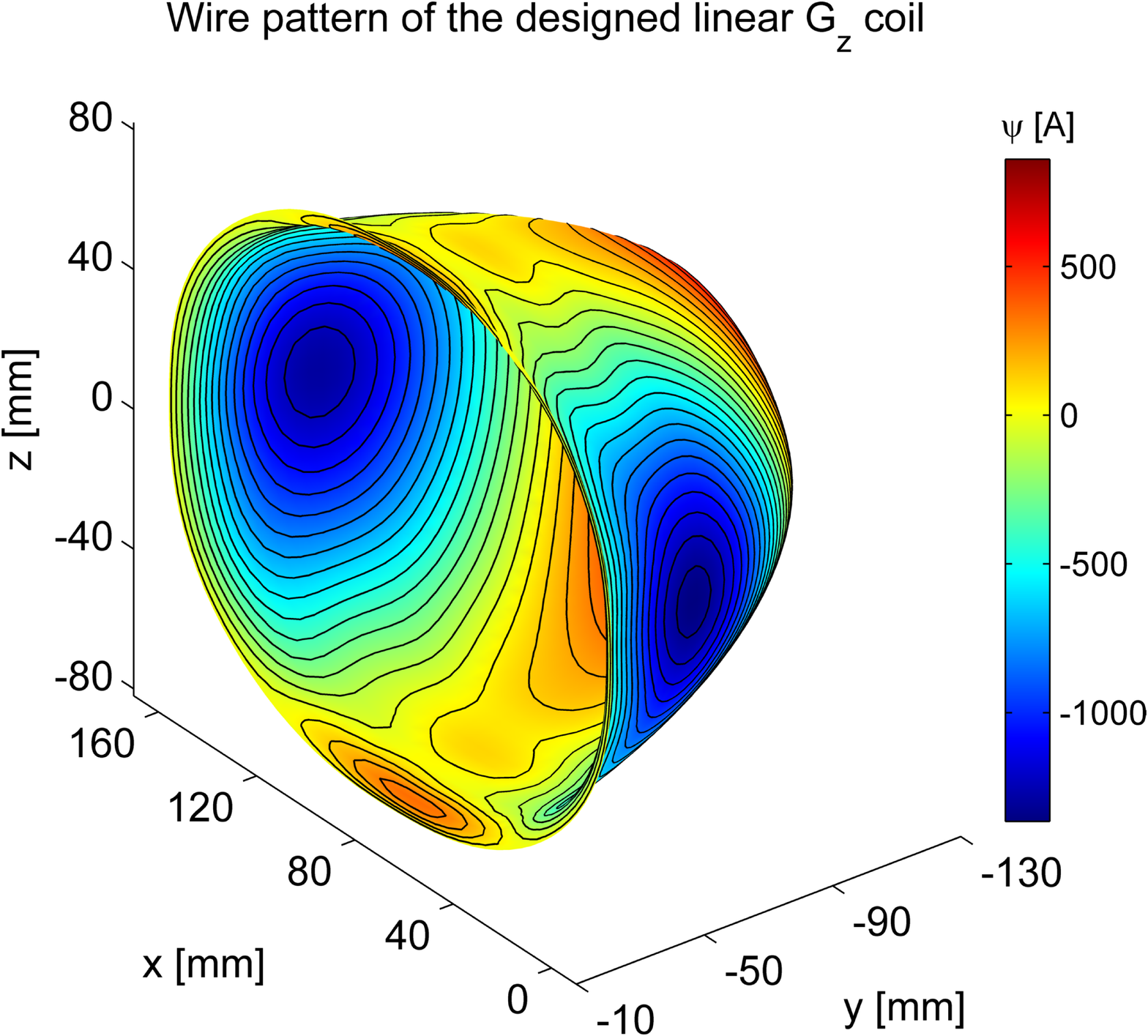}
	}\caption{Coil layouts of the non-linear coils designed without (a) or with (b) constraint ($\ref{equ:NBCD_CoronalSliceConstraints}$) and the linear $G_z$ coil (c).}
	\label{fig:SFCoilLayoutNoBrim_fig3} 
\end{figure}

Figure \ref{fig:SFCoilLayoutNoBrim_fig3} shows wire patterns of the two non-linear coils and the linear $G_z$ coil on $\Gamma$. As seen, each of the three coils contains some wires that touch the outer boundary of the current-carrying surface $\Gamma$. These wires are important to generate high gradient strength of the encoding field $B_z$ within coronal slices of the ROI close to the chest. Unfortunately, as depiced in Figure \ref{fig:EtaXfemNoBrim_fig2}.a-b, the two non-linear coils produce lower gradient strengths in these slices than other areas within the ROI. However, breast regions covered by these coronal slices need also be well depicted by DWI for cancer diagnosis [ref]. Therefore, it is desirable to design a non-linear coil, able to generate higher gradient strengths in these slices, than the two non-linear coils presented above. One possible solution is to extend the current-carrying surface $\Gamma$ with a rim. On this extended surface $\Gamma_r$, higher current density can be placed near to the proximal slices and may help to generate higher gradient strength there. The detailed coil design on $\Gamma_r$ will be presented in the following subsection.
 
\subsection{Coil design for a current-carrying surface with a rim}
\begin{figure}[htbp]
	\centering \subfigure[]{ \label{fig:SFCoilLayout1_fig4a} \includegraphics[width=0.31\textwidth]{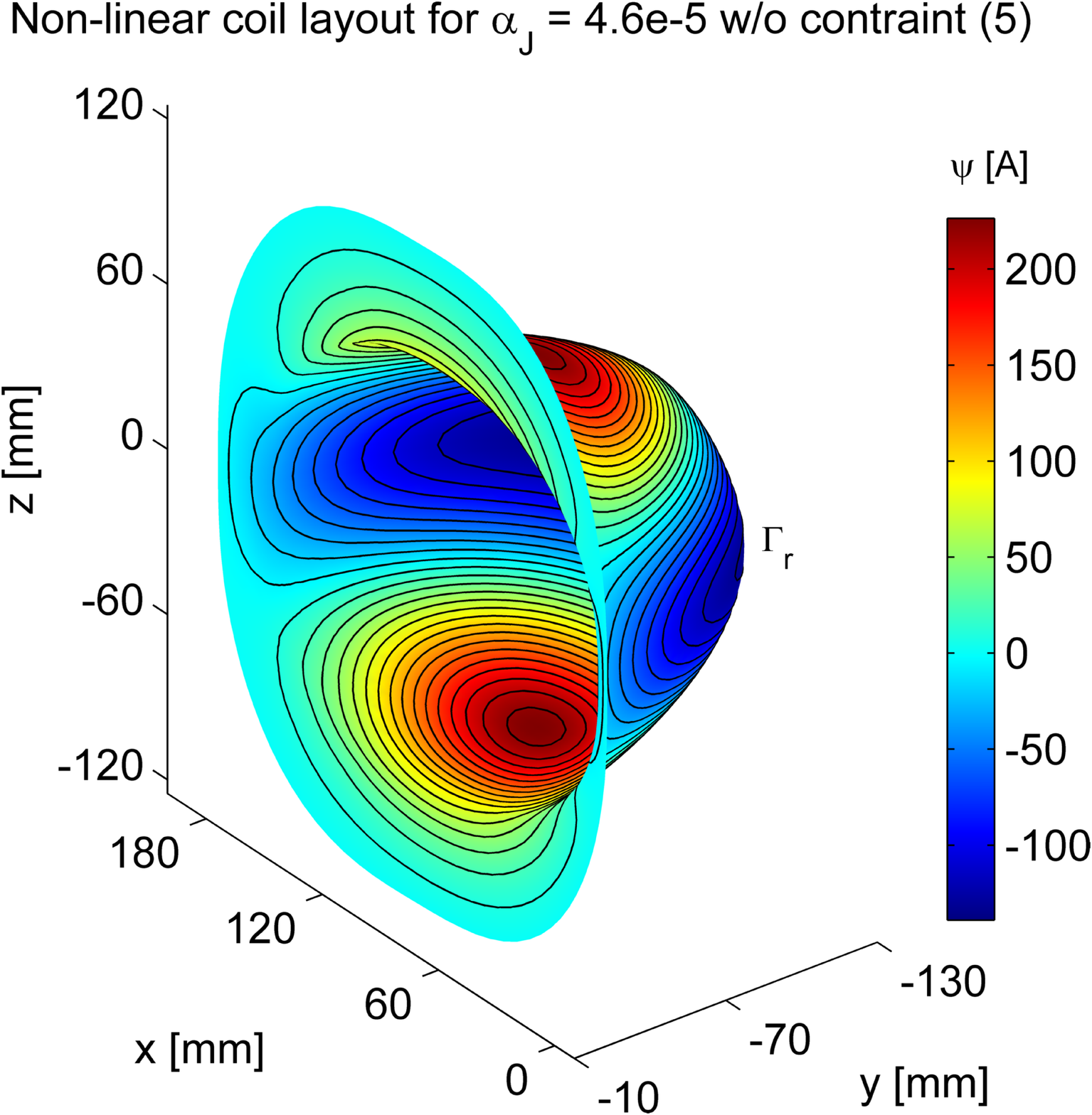}
	} \subfigure[]{ \label{fig:SFCoilLayout2_fig4b} \includegraphics[width=0.31\textwidth]{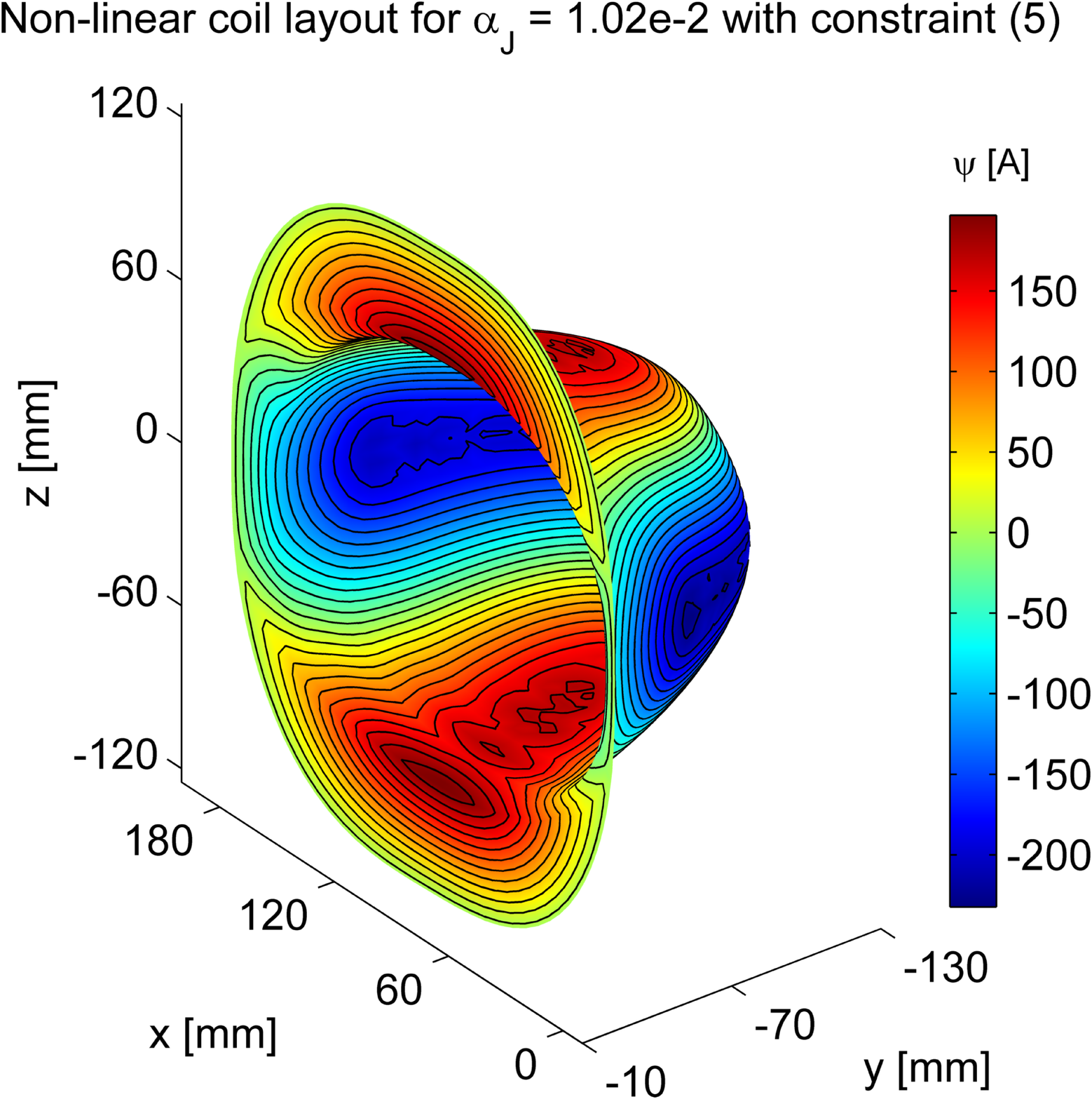}
	} \subfigure[]{ \label{fig:SFCoilLayout3_fig4c} \includegraphics[width=0.31\textwidth]{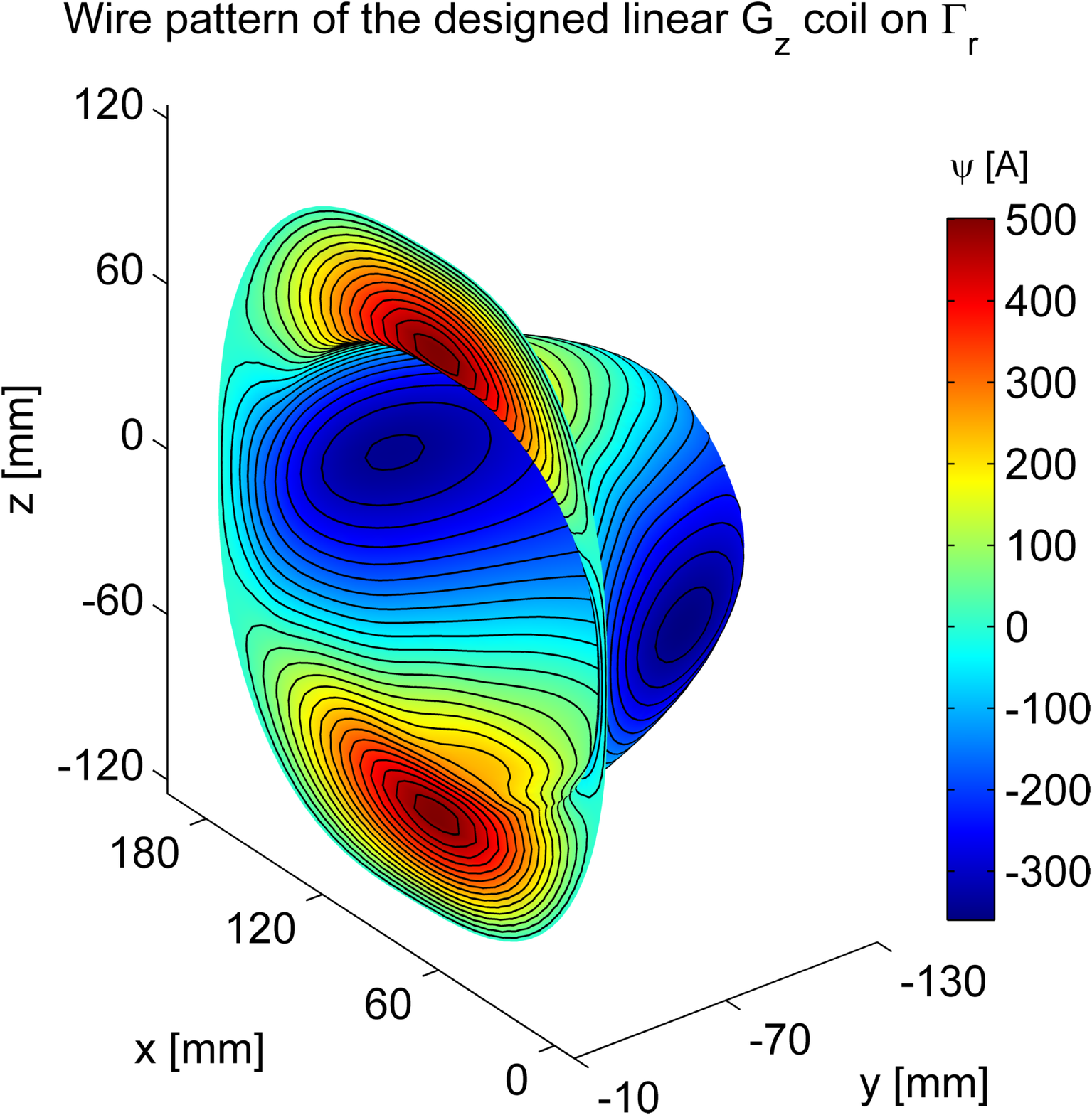}
	}\caption{Coil layouts of the non-linear coils designed on the extended surface $\Gamma_r$ without (a) or with (b) constraint ($\ref{equ:NBCD_CoronalSliceConstraints}$) and the linear $G_z$ coil (c).}
	\label{fig:SFCoilLayout_fig4} 
\end{figure}

Figure \ref{fig:SFCoilLayout_fig4} shows wire patterns of the two non-linear coils and the linear $G_z$ coil designed on the extended current-carrying surface $\Gamma_r$. As seen, compared to one non-linear coil without constraint (\ref{equ:NBCD_CoronalSliceConstraints}) on $\Gamma_r$, the other non-linear coil and the linear coil contain more winding wires in the rim part of $\Gamma_r$. This phenomenon indicates that the extended rim part of $\Gamma_r$ is more useful for the designs of the linear coil and the non-linear coil with constraint (\ref{equ:NBCD_CoronalSliceConstraints}) than the non-linear coil without the constraint.

\begin{figure}[htbp]
	\centering \subfigure[]{ \label{fig:EtaXfem1_fig5a} \includegraphics[width=0.425\textwidth]{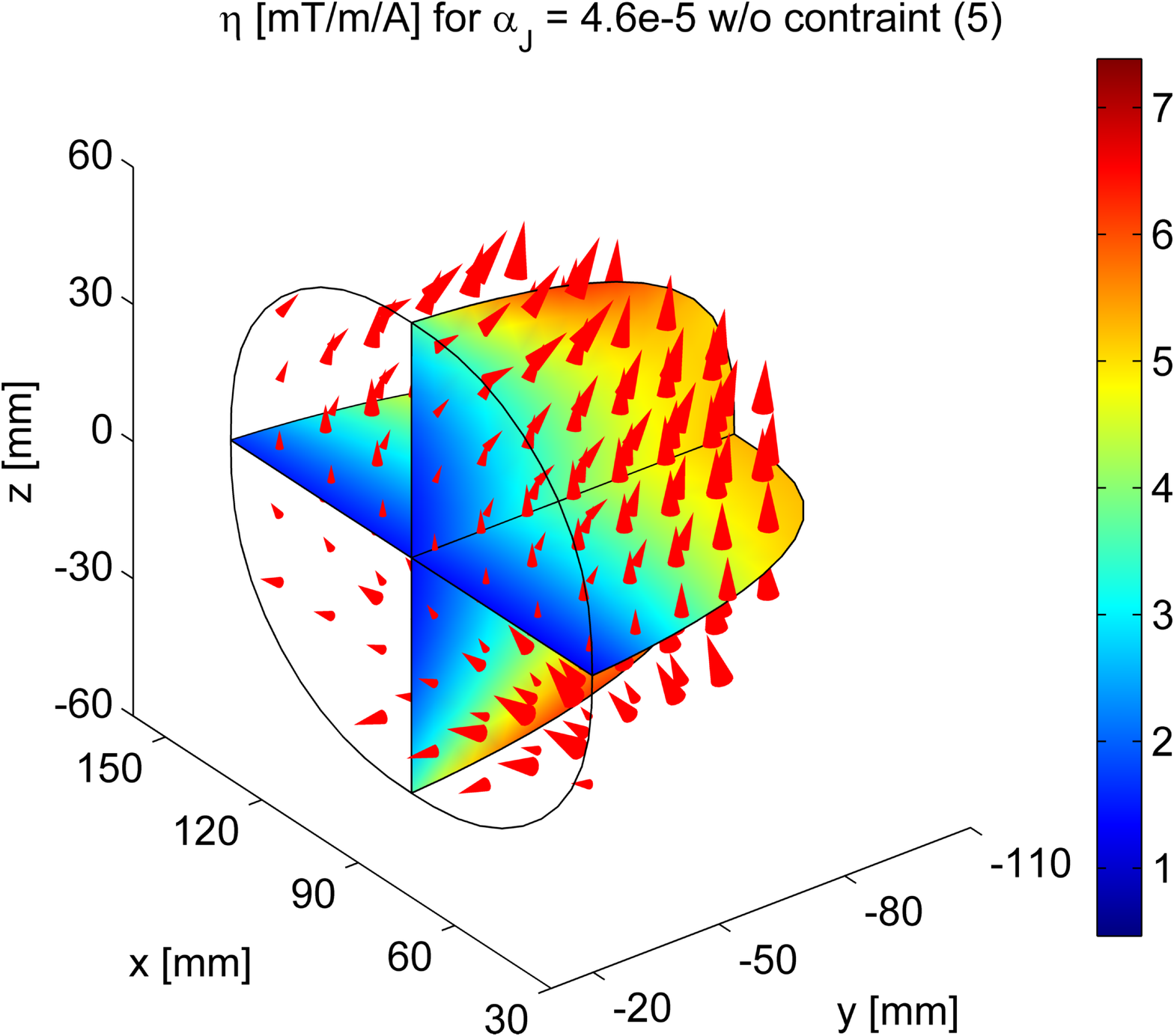}
	} \subfigure[]{ \label{fig:EtaXfem2_fig5b} \includegraphics[width=0.425\textwidth]{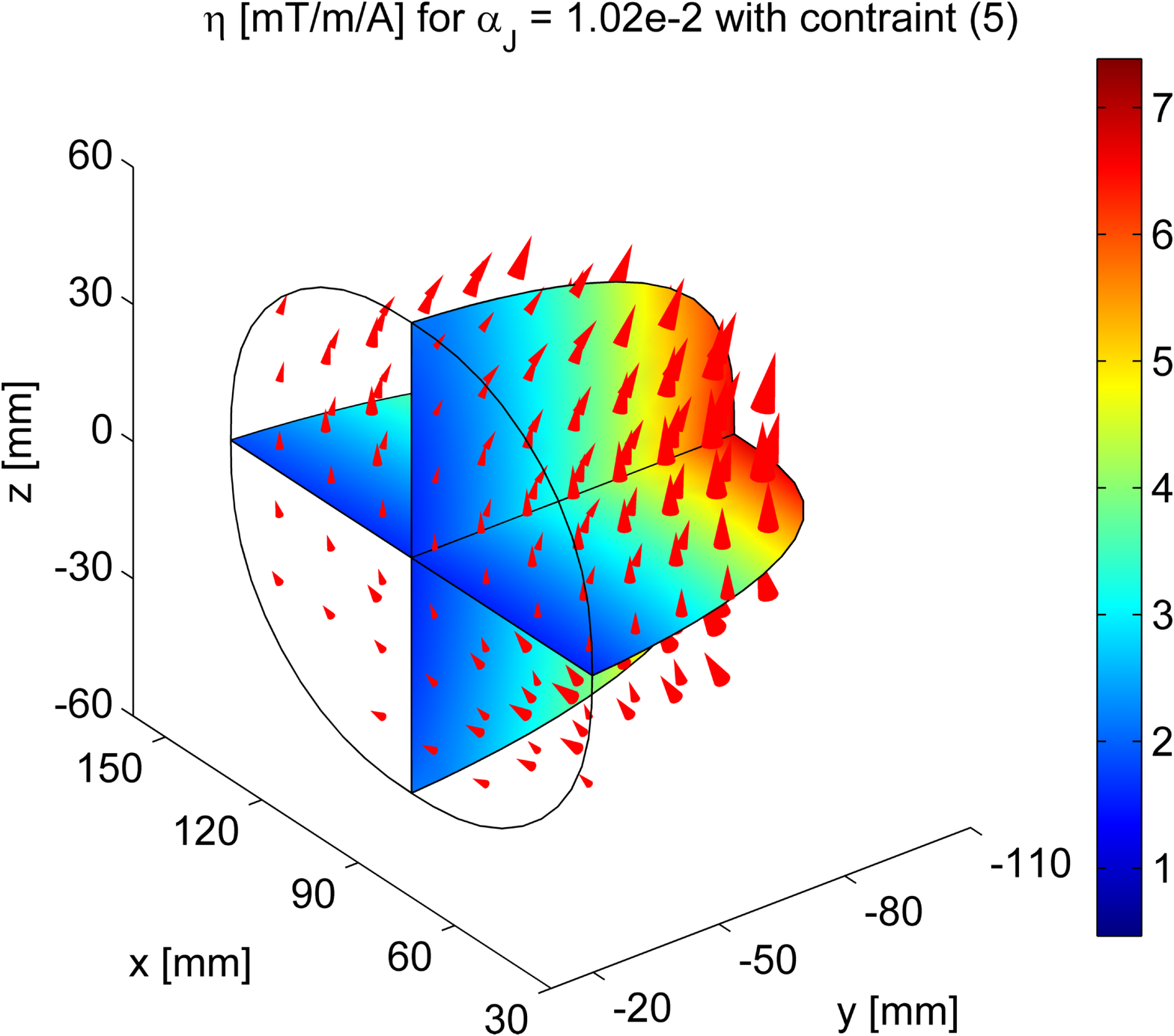}
	} \subfigure[]{ \label{fig:EtaXfem3_fig5c} \includegraphics[width=0.425\textwidth]{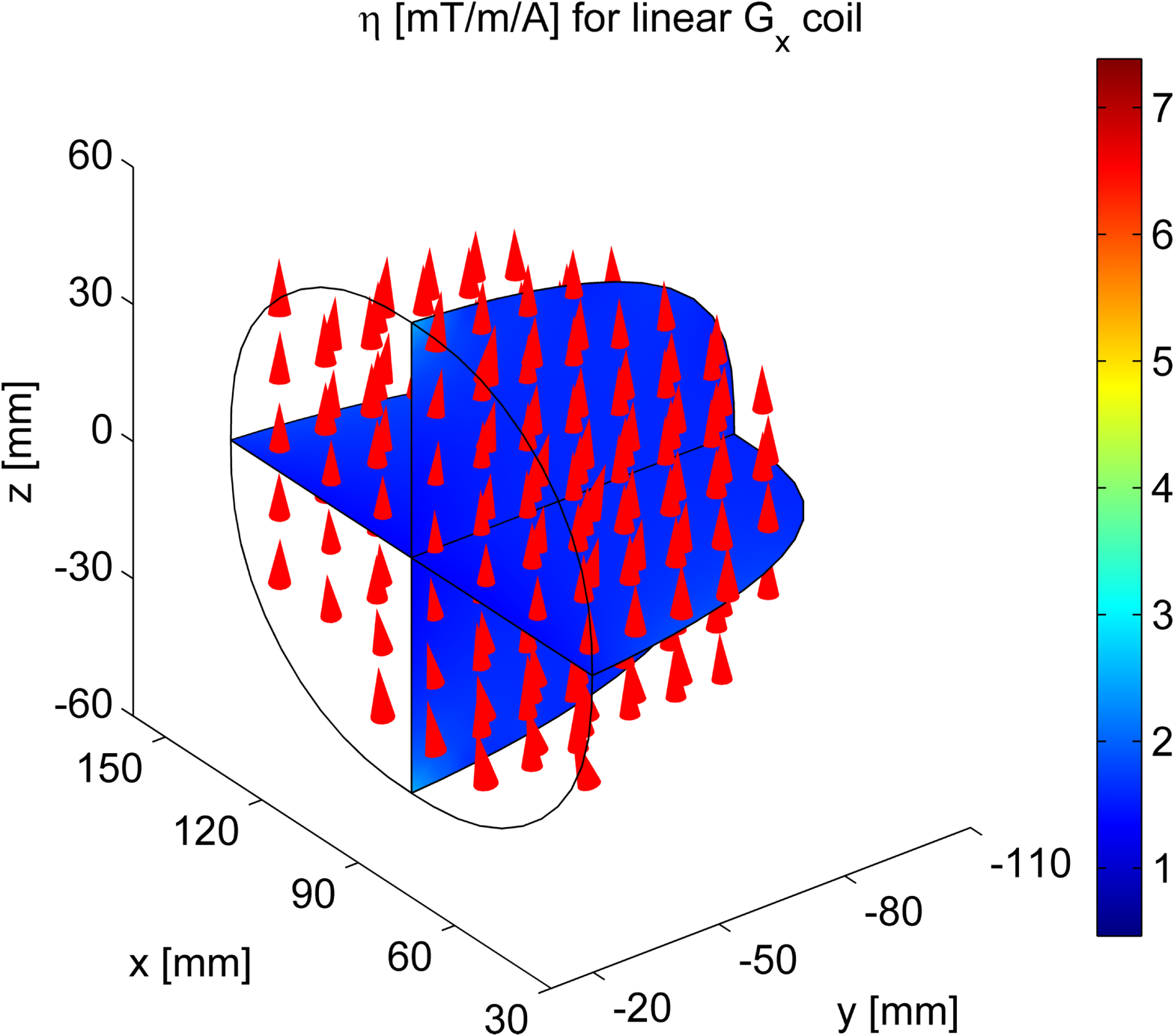}
	} \subfigure[]{ \label{fig:HistCompROI_fig5d} \includegraphics[width=0.425\textwidth]{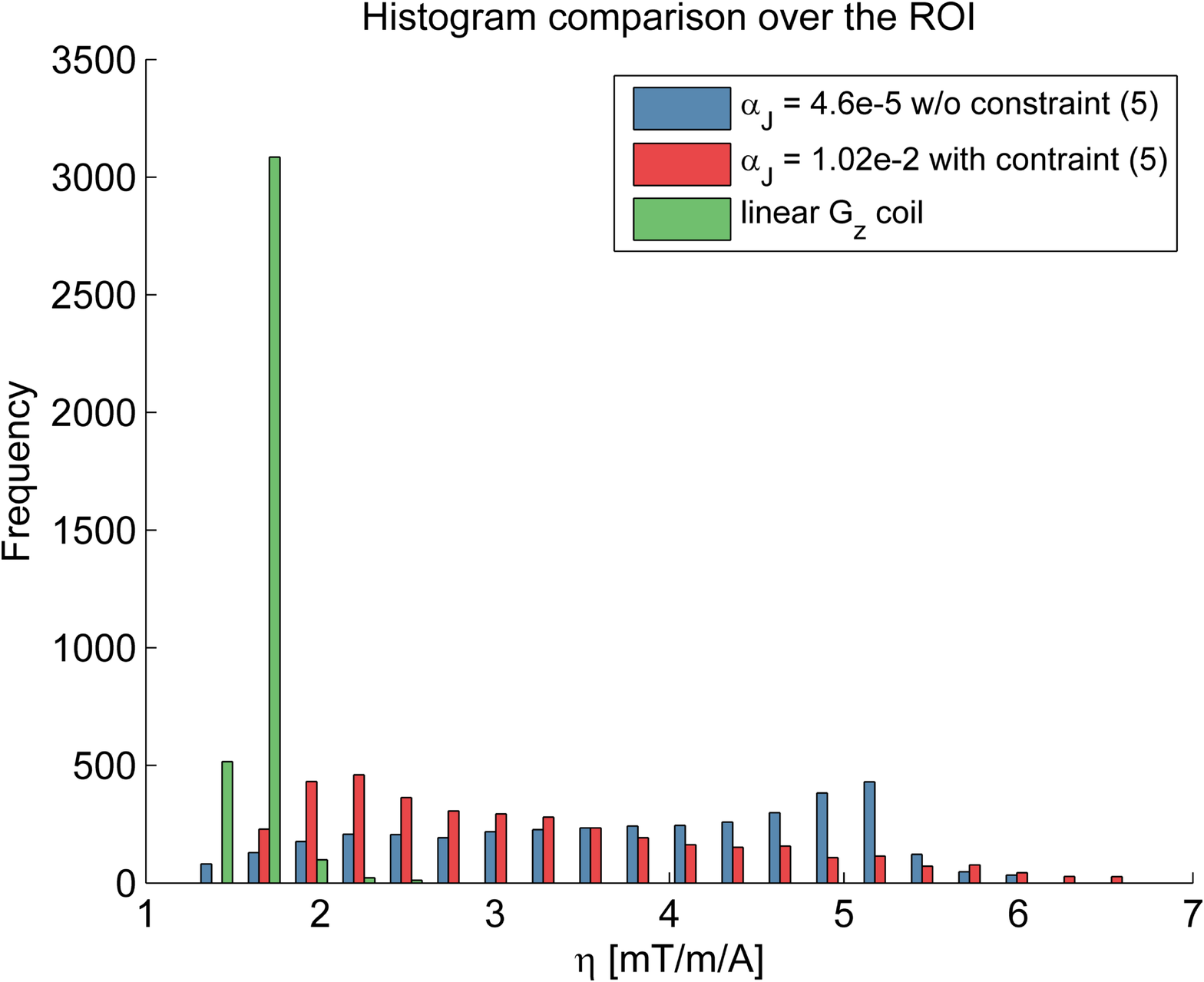}
	} \subfigure[]{ \label{fig:HistCompS1_fig5e} \includegraphics[width=0.425\textwidth]{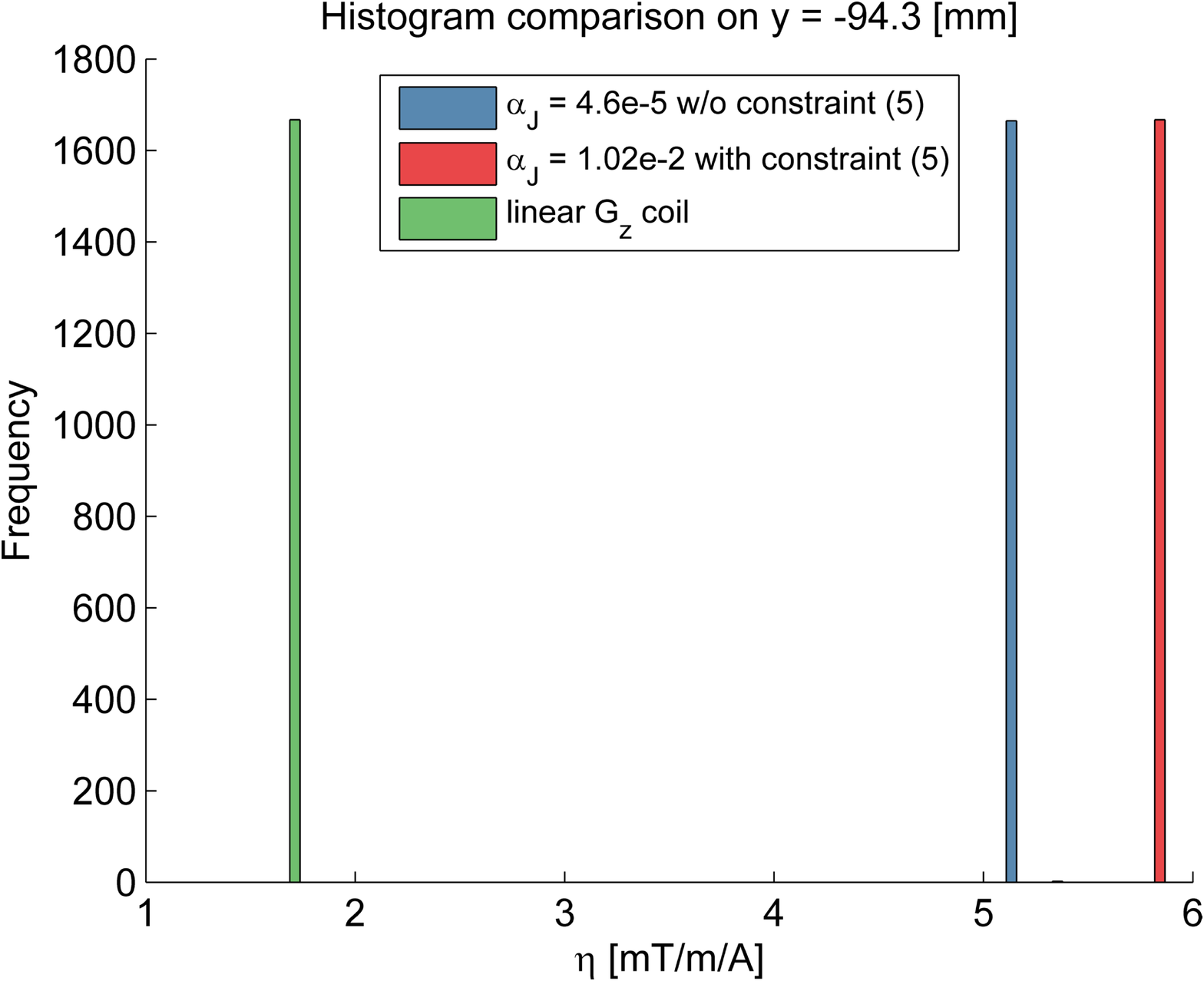}
	} \subfigure[]{ \label{fig:HistCompS5_fig5f} \includegraphics[width=0.425\textwidth]{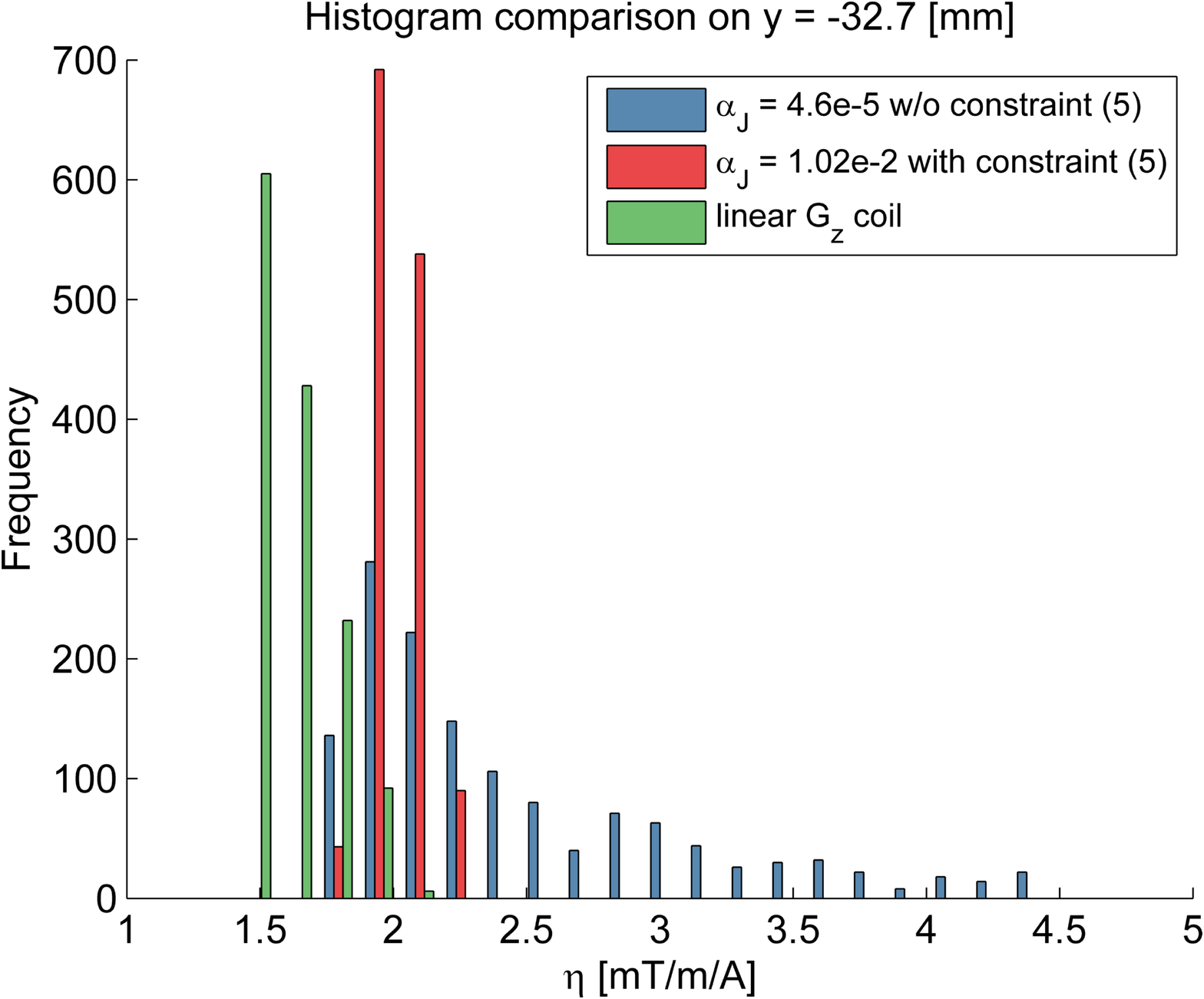}
	}\caption{Coil efficiency $\eta$ for the non-linear coils without (a) and with (b) constraint ($\ref{equ:NBCD_CoronalSliceConstraints}$) and the linear $G_z$ coil (c) designed on the extended surface $\Gamma_r$. Here, gradient vectors of the $B_z$ field generated by the three coils with one unit current are marked in red arrows and the arrow lengths are proportional to the magnitude of the gradient vectors. The histogram (d) presents a rough probability distribution of coil efficiency $\eta$ over the ROI for these three coils. The histograms (e) and (f) show the distribution of $\eta$ within two representative coronal slices of $y=-94.3$ mm and $y=-32.7$ mm, respectively.}
	\label{fig:EtaXfem_fig5} 
\end{figure}

Figure \ref{fig:EtaXfem_fig5} presents the distribution of coil efficiency $\eta$ for these three coils designed on the extended surface $\Gamma_r$ and Table \ref{table:NBCD_Non-LinearCoilsGammaR} displays properties of these coils. As depicted in Figures \ref{fig:EtaXfem_fig5}a-c and \ref{fig:EtaXfemNoBrim_fig2}a-c, minimum of coil efficiency $\eta$ over the ROI for these three coils have higher values as compared to the corresponding designed coils on the surface $\Gamma$. Notably, as seen in Tables~\ref{table:NBCD_LinearCoils}, \ref{table:NBCD_Non-LinearCoils} and~\ref{table:NBCD_Non-LinearCoilsGammaR}, minima of $\eta$ for the linear $G_z$ coil, the non-linear coil with and without Constraint (\ref{equ:NBCD_CoronalSliceConstraints}) on $\Gamma_r$ increase by 2.84, 2.21 and 1.15 times than that for the corresponding coils on $\Gamma$, respectively. Moreover, the linear $G_z$ coil, the non-linear coil with and without Constraint (\ref{equ:NBCD_CoronalSliceConstraints}) on $\Gamma_r$ show 2.70, 1.17 and 1.04 fold increase in the resistive figure of merit $\beta_P$ than their counterparts on $\Gamma$, respectively. Although the inductive figure of merit $\beta_W$ for the non-linear coil without Constraint (\ref{equ:NBCD_CoronalSliceConstraints}) on $\Gamma_r$ reduces by a factor of 1.01 as compared to the corresponding coil on $\Gamma$, $\beta_W$ for the linear coil and the other non-linear coil on $\Gamma_r$ grow 2.60- and 1.11-fold, respectively. These results suggest that the additional rim in $\Gamma_r$ has a relatively small influence on the performance in a non-linear coil design without Constraint (\ref{equ:NBCD_CoronalSliceConstraints}) while making a large impact both on the linear coil design and non-linear coil design with Constraint (\ref{equ:NBCD_CoronalSliceConstraints}). This observation is compatible with the wire patterns of these three coils, as shown in Figure \ref{fig:SFCoilLayout_fig4}.

Although the two figures of merit for the linear $G_z$ coil show more than 2.5-fold increase upon the extension of the surface $\Gamma$ to $\Gamma_r$, $\beta_P$ and $\beta_W$ for the non-linear coil without Constraint (\ref{equ:NBCD_CoronalSliceConstraints}) increase by around 2.39 and 2.12 times, respectively (Table \ref{table:NBCD_Non-LinearCoilsGammaR}). For the non-linear coil with the constraint, these parameters grow by 1.80- and 1.69-fold, respectively. Moreover, Table \ref{table:NBCD_Non-LinearCoilsGammaR} also reveals that the maximum of coil efficiency $\eta$ for the non-linear coils without and with Constraint (\ref{equ:NBCD_CoronalSliceConstraints}) more than 2.46 and 2.68 times higher than that for the linear coil. However, as shown in Figure \ref{fig:EtaXfem_fig5}.a for a non-linear coil without Constraint (\ref{equ:NBCD_CoronalSliceConstraints}), three-dimensional curved isosurfaces of $\eta$ with a high curvature can be observed in the areas proximal to the chest, possibly posing a challenge to a data acquisition and image reconstruction in DWI.

Unlike the non-linear coil on $\Gamma_r$ without Constraint (\ref{equ:NBCD_CoronalSliceConstraints}), $\eta$ for the non-linear coil with the constraint tends to remain constant in every coronal slice (Figure \ref{fig:EtaXfem_fig5}.b), possibly overcoming the challenge to data acquisition on a coronal slice. For example, as illustrated in Figure \ref{fig:EtaXfem_fig5}.e and \ref{fig:EtaXfem_fig5}.f, $\eta$ within two representative coronal slices at $y=-94.3$ mm and $y=-32.7$ mm for the linear coil and the non-linear coil with constraint (\ref{equ:NBCD_CoronalSliceConstraints}) have distributions with a lower dispersion than the non-linear coil without the constraint. Particularly, the ratios of the standard deviation to the mean of $\eta$ within the coronal slice at $y=-32.7$ mm are 0.0496 and 0.0946 for the linear coil and the non-linear coil with constraint (\ref{equ:NBCD_CoronalSliceConstraints}), respectively. However, this ratio in the same slice for the non-linear coil without the constraint increases by 2.72 and 5.19 times compared to these two coils, respectively.

Figures \ref{fig:EtaXfem_fig5}.e and \ref{fig:EtaXfem_fig5}.f also reveal that the non-linear coil with Constraint (\ref{equ:NBCD_CoronalSliceConstraints}) on $\Gamma_r$ has a much higher mean coil efficiency $\eta$ within coronal slices than the linear coil. For example, the mean of $\eta$ for the non-linear coil increases more than 3.78-fold in the slice at $y=-94.3$ mm. For example, to generate the gradient strength of 1 T/m in this slice, the non-linear coil requires a current of less than 168 A and will dissipate 715 W of power. To obtain the same gradient strength using the linear coil, more than 3.78 times the current and an 11.45-fold increase in dissipated power would be needed.

\section{Discussion}
\label{sec.Discussion}
In this study, we have developed two novel approaches to design high-performance non-linear gradient coils for diffusion weighting imaging in the female breast. The first approach is similar to the method used to design elements of a matrix coil~\citep{jia_design_2017,LittinZaitsev2017a}. The main difference is that coil dissipated power is minimized in the first approach while mean gradient strength of the non-linear encoding field was maximized in the previous coil element design. This change of objectives enables us to easily obtain a designed non-linear coil comparable to a corresponding linear coil which is a solution of Problem (\ref{equ:NBCD_LinearP}). Moreover, force and torque balance are also considered in the current approaches. The second approach has been developed by extending the first method with a constraint of an encoding field variation in coronal slices. This constraint can help to address the challenge on data acquisition and image reconstruction for DWI, which otherwise is expected to arise when using non-linear gradient coils for diffusion weighting. Numerical results have demonstrated the effectiveness of these two approaches. Notably, an ultra-strong gradient of above 1T/m in a local region of the ROI can be achieved using the designed non-linear coils with a current of less than 170 A. After designing coils on two current-carrying surfaces with different approaches, three figures of merit $\beta_P$, $\beta_W$ and $\nu_P$ were used to assess the performance of different coils. 

Tables~\ref{table:NBCD_Non-LinearCoils} and~\ref{table:NBCD_Non-LinearCoilsGammaR} show that inductances of the designed non-linear breast coils are below 100 $\mu$H with the minimum wire width limited to at least 2.5 mm. Here, a Litz wire with the width of at least 2.2~mm and the thickness of 5.8 mm (Rudolf Pack GmbH, Gummersbach, Germany) can be used for coil fabrication. This ensures that the coils built based on the presented designs can support driving currents of about 170~A, thus generating gradient strengths of above 1~T/m in practice. These results also demonstrate that the assumption of a low coil inductance introduced in the section \ref{sec:Introduction} is reasonable for the considered coil designs.

In the second approach as described in Problem (\ref{equ:NBCD_non-linearF}), the norm of the two-dimensional gradient of the gradient amplitude of the encoding field within each of the coronal slices is controlled using Constraint (\ref{equ:NBCD_CoronalSliceConstraints}). The coronal orientation of the slices was found to be advantageous due to the geometry of the current carrying surface, expected to allowing maximization of the diagnostic image quality in a clinical protocol. Indeed, since there is no possibility to allocate any current density to the open side of the cup, there is a natural decline of the achievable gradient strength in the vertical direction (corresponding to the y axis of the MR system). Enforcing Constraint (\ref{equ:NBCD_CoronalSliceConstraints}) for transversal or sagittal slices would attempt to suppress this natural variation leading to a substantial drop in the maximum achievable gradient strengths. It is worth noting that our proposed approach can be applied to design high-performance non-linear gradient coils for other human organs, such as prostate and brain, but would possibly require a different preferred slice orientation. Further analysis of the related problems and possible solutions is a topic of our future research work.

Figures \ref{fig:SFCoilLayoutNoBrim_fig3} and \ref{fig:EtaXfem_fig5} indicate that a suitable selection of $C_c$ in Constraint (\ref{equ:NBCD_CoronalSliceConstraints}) is effective for reducing inhomogeneities of the gradient amplitude of the resulting encoding fields in each of the coronal slices. However, it is generally unclear, which values of the $C_c$ should be considered reasonable. In the present paper the approach was taken to estimate $C_c$ based on the encoding field generated by the designed linear $G_z$ coil. The main rationale behind such a procedure for selecting $C_c$ was that the gradient linearity of conventional designs with a similar tolerance (5\% field error) is generally considered as acceptable or at least manageable. As demonstrated by other researchers for diffusion weighted imaging~\cite{bammer_analysis_2003, tan_improved_2013}, effects of gradient non-linearities generated by linear gradient coils including their concomitant fields can be effectively handled by appropriate correction approaches. This means that with the current selection of $C_c$ the present correction approaches for reducing the possible effects of gradient non-linearities are likely to be applicable. Further research will show if any such corrections may be required for our target application.

Tables~\ref{table:NBCD_Non-LinearCoils} and~\ref{table:NBCD_Non-LinearCoilsGammaR} also indicate that
all coils designed on the extended surface $\Gamma_r$ with a rim outperform those on $\Gamma$ in terms of both resistive and inductive figures of merit. However, many windings of these coils are located on the rim, which is close to the chest. Therefore, the coils designed on $\Gamma_r$ may generate higher electrical field inside the chest, thus possibly leading to a higher peripheral nerve stimulation (PNS). In order to use the coils on $\Gamma_r$ in vivo, effects of PNS need to be investigated. One possible solution to predict the PNS threshold of the designed coils is to perform electromagnetic field simulation including realistic body models proposed by Davids et al.~\cite{davids_predicting_2017}. Another possible solution is to build prototype coils and assess the PNS effects experimentally. Both of these routes are ongoing work at our institution.

This study only focuses on unshielded non-linear coil designs for the following two reasons. First, the purpose of this study is to design a non-linear local coil to obtain as high a gradient strength as possible for a given power dissipation. It is well known that unshielded coils have a high efficiency. Second, as the designed local coil is far from the cryostat, it is expected to generate only minor eddy currents. Practical effects of induced eddy currents induced by 
the coils will be assessed experimentally after a prototype of the non-linear coil is realized. If the observed eddy current effects due to diffusion weighted imaging turn out to be substantial, two possible approaches will be explored to reduce the effects in future work.  One is to design an actively-shielded non-linear breast coil; the other is to use conventional linear gradient coil systems to compensate the effects as described in~\cite{van_der_velden_novel_2017}. Undoubtedly a system allowing for the examination of both breasts simultaneously is of advantage in a diagnostic setting when searching for suspicious lesions. However, when aiming to provide additional information about an already detected and located unclear lesion, a single-side coil system as described herein might find its place in expanding the diagnostic toolbox of radiologists, reducing the need for invasive procedures and gadolinium exposure for the patient.

\section{Conclusion}
\label{sec:Conclusion}
Gradient coil design for an irregularly shaped semi-open current-carrying surface with a non-spherical target region has been considered. The multi-objective optimization problem was reformulated as a constrained optimization, allowing for flexible and precise control of the coil properties. A novel constraint was proposed limiting the gradient magnitude variation within each slice while allowing for variations in both the direction of the gradient within the slice and the magnitude across the slices. These innovations allowed us to design a unilateral coil for diffusion weighting in the female breast with the local gradient strengths exceeding 1~T/m with a high homogeneity of the diffusion weighting for imaging in the coronal slice orientation. 


\section*{Acknowledgments}
This work was supported by the German Research Foundation (DFG) (Grant Number ZA 422/5-1 and ZA 422/6-1). 

\bibliography{NBCD}
\bibliographystyle{elsarticle-num}

\newpage{}

\paragraph{Figure captions}

Figure 1. Geometries of the ROI and current-carrying surface $\Gamma$. The detailed parameters are presented in Table \ref{table:NBCD_LinearCoils}.

Figure 2. Coil efficiency $\eta$ for the designed non-linear coils without (a) and with (b) constraint ($\ref{equ:NBCD_CoronalSliceConstraints}$). In order to reveal high performance of the non-linear coils, coil efficiency $\eta$ for the corresponding linear $G_z$ coil (c) is also plotted. Here, Here, gradient vectors of the $B_z$ field generated by the three coils with one unit current are marked in red arrows and the arrow lengths are proportional to the magnitude of the gradient vectors. The histogram (d) presents a rough probability distribution of coil efficiency $\eta$ over the ROI for these three coils. The histograms (e) and (f) show the distribution of $\eta$ within two representative coronal slices of $y=-94.3$ mm and $y=-32.7$ mm, respectively.

Figure 3. Coil layouts of the non-linear coils designed without (a) or with (b) constraint ($\ref{equ:NBCD_CoronalSliceConstraints}$) and the linear $G_z$ coil (c).

Figure 4. Coil layouts of the non-linear coils designed on the extended surface $\Gamma_r$ without (a) or with (b) constraint ($\ref{equ:NBCD_CoronalSliceConstraints}$) and the linear $G_z$ coil (c).

Figure 5. Coil efficiency $\eta$ for the non-linear coils without (a) and with (b) constraint ($\ref{equ:NBCD_CoronalSliceConstraints}$) and the linear $G_z$ coil (c) designed on the extended surface $\Gamma_r$. Here, gradient vectors of the $B_z$ field generated by the three coils with one unit current are marked in red arrows and the arrow lengths are proportional to the magnitude of the gradient vectors. The histogram (d) presents a rough probability distribution of coil efficiency $\eta$ over the ROI for these three coils. The histograms (e) and (f) show the distribution of $\eta$ within two representative coronal slices of $y=-94.3$ mm and $y=-32.7$ mm, respectively.

\newpage{}

\begin{table}[htb]
\centering
\begin{tabular}{c|ccc}
\hline Properties  & Linear $G_x$ coil & Linear $G_y$ coil & Linear $G_z$ coil \\ 
\hline $\left[\min \max\right]$ aperture in x,z [mm] & \multicolumn{3}{c}{[-2.39 168.65], [-80.78 80.786]} \\
$\left[\min \max \right]$ depth in y [mm] & \multicolumn{3}{c}{[-126.5 -19.97]} \\
ROI-x-y-z [mm] &  \multicolumn{3}{c}{112.4-77-102.88} \\
Target field $B_z^*$ [mT] & 50.0(x-mean$(x_i)$) & 50.0(y-mean$(y_i)$) & 50.0z \\
 Field accuracy $\delta$ [\%] & 5 & 5 & 5 \\
 Number of contours & 26 & 26  & 26 \\
 Current $I$ [A]& 195.68 & 3.741e3  & 85.6 \\
 $\eta:=|\nabla{B_z}|/I$ [mT/m/A] & [0.185,0.4725] & [3.93e-3,39.14e-3]& [0.4654,1.0] \\ 
 $\beta_P$ [T/(m$\cdot$A$\cdot\Omega^{1/2}$)] & 7.5685e-4 & 1.767e-5 & 4.2$\times10^{-3}$\\
 $\beta_W$ [T/(m$\cdot$A$\cdot H^{1/2}$)] & 0.0571 & 0.0016 & 0.124\\
 $\nu_P$ [T/(m$^2\cdot$A$\cdot\Omega^{1/2}$)] & 2.74e-4 & 2.97e-5 & 7.017e-4\\
 dissipated power $P$ [W] & 4.986e3 & 8.29e6 & 158.4 \\ 
 Magnetic energy $W$ [J] & 0.8775 & 1.0124e3 & 0.184 \\ 
 Coil inductance [$\mu$H] & 45.83 & 144.67 & 50.24 \\
 Coil resistance [m$\Omega$] & 130.2 & 592.6 & 21.6 \\
 $\max (|M_x|,|M_y|,|M_z|)$  [N$\cdot$m] & 1.0e-4 & 1.0e-4 & 1.0e-4 \\
\hline 
\end{tabular} 
\caption{Geometric parameters of the current-carrying surface $\Gamma$, the ROI (Fig. \ref{fig:CCSMeshROI_fig1}) and performance comparison among linear gradient coils.} \vspace{2mm}
\label{table:NBCD_LinearCoils}
\end{table}

\begin{table}[htb]
	\centering
	\begin{tabular}{c|c|c}
		\hline Properties  & \tabincell{c}{Non-linear coil \\ w/o constraint (\ref{equ:NBCD_CoronalSliceConstraints}) \\ in coronal slices}& \tabincell{c}{Non-linear coil \\ with constraint (\ref{equ:NBCD_CoronalSliceConstraints})\\ in coronal slices} \\ 
		\hline $\alpha_J$ & 1.94e-4 & 3.4e-4  \\
		$C_g$ & 0.0534 & 0.0534  \\
		$C_c$ & - & 7.017$\times10^{-4}$  \\
		Number of contours & 26 & 26   \\
		Current $I$ [A]& 13.73 & 22.87   \\
		$\eta:=|\nabla{B_z}|/I$ [mT/m/A] & [1.1,6.7] & [0.702,7.4]  \\ 
		$\beta_P$ [T/(m$\cdot$A$\cdot\Omega^{1/2}$)] & 0.026 & 0.0174 \\
		$\beta_W$ [T/(m$\cdot$A$\cdot H^{1/2}$)] & 0.6906 & 0.4932 \\
		$\nu_P$ [T/(m$^2\cdot$A$\cdot\Omega^{1/2}$)] & 6.875e-3 & 7.017e-4 \\
		dissipated power $P$ [W] & 4.2 & 9.42  \\ 
		Magnetic energy $W$ [J] & 4.2e-3 & 0.0117  \\ 
		Coil inductance [$\mu$H] & 63.4 & 44.8  \\
		Coil resistance [m$\Omega$] & 22.3 & 18  \\
 $\max (|M_x|,|M_y|,|M_z|)$  [N$\cdot$m] & 1.0e-4 & 1.0e-4 \\
		Mimimum wire width [mm] & 2.5 & 2.5  \\
		\hline 
	\end{tabular} 
	\caption{Performance comparison among non-linear gradient coils designed on the current-carrying surface $\Gamma$ shown in Fig. 1.} \vspace{2mm}
	\label{table:NBCD_Non-LinearCoils}
\end{table}

\begin{table}[htb]
	\centering
	\begin{tabular}{c|ccc}
		\hline Properties  & \tabincell{c}{Non-linear coil \\ w/o constraint (\ref{equ:NBCD_CoronalSliceConstraints})}& \tabincell{c}{Non-linear coil \\ with constraint (\ref{equ:NBCD_CoronalSliceConstraints})} & Linear $G_z$ coil \\ 
		\hline $\left[\min \max\right]$ aperture in x,z [mm] & \multicolumn{3}{c}{[-5 185], [-120 120]} \\
		$\left[\min \max \right]$ depth in y [mm] & \multicolumn{3}{c}{[-126.5 -18.54]} \\
		ROI-x-y-z [mm] &  \multicolumn{3}{c}{112.4-77-102.88} \\
		$\alpha_J$ & 4.6e-5 & 1.02e-4 & - \\
		$C_g$ & 0.0534 & 0.0534 & - \\
		$C_c$ & - & 7.017e-4 & - \\
		Number of contours & 26 & 26  & 26 \\
		Current $I$ [A]& 14.05 & 16.51  & 33.14 \\
		$\eta:=|\nabla{B_z}|/I$ [mT/m/A] & [1.27,6.15] & [1.55,6.7] & [1.323,2.497] \\ 
		$\beta_P$ [T/(m$\cdot$A$\cdot\Omega^{1/2}$)] & 0.027 &0.0203 & 0.0113\\
		$\beta_W$ [T/(m$\cdot$A$\cdot H^{1/2}$)] & 0.683 & 0.5452 & 0.3227\\
		$\nu_P$ [T/(m$^2\cdot$A$\cdot\Omega^{1/2}$)] & 6.336e-3 & 7.017e-4 & 1.398e-3\\
		dissipated power $P$ [W] & 3.895 & 6.926 & 22.32 \\ 
		Magnetic energy $W$ [J] & 6.1e-3 & 9.6e-3 & 0.0274 \\ 
		Coil inductance [$\mu$H] & 61.85 & 70.31 & 49.96 \\
		Coil resistance [m$\Omega$] & 19.7 & 25.4 & 20.3 \\
		 $\max (|M_x|,|M_y|,|M_z|)$  [N$\cdot$m] & 1.0e-4 & 1.0e-4 & 1.0e-4 \\
		Mimimum wire width [mm] & 2.5 & 2.5 & 1.57 \\
		\hline 
	\end{tabular} 
	\caption{Geometric parameters of the extended current-carrying surface $\Gamma_r$ (Fig. \ref{fig:SFCoilLayout_fig4}.a) and performance comparison among the coils designed on the $\Gamma_r$. Here, The same ROI, as shown in Fig. \ref{fig:CCSMeshROI_fig1}, was used.}\vspace{2mm}
	\label{table:NBCD_Non-LinearCoilsGammaR}
\end{table}

\newpage{}

\newpage{}

\end{document}